\documentstyle[12pt,aasms4]{article}
\begin{document}
\title{Multiwaveband Observations of Quasars with Flat Radio Spectra and
Strong Millimeter Wave Emission}
\author{Steven D. Bloom \altaffilmark{1,}\altaffilmark{2}, Alan P. Marscher, E. M. Moore \altaffilmark{3}} 
\affil{Department of Astronomy, Boston University, 725 Commonwealth Avenue,
Boston, MA 02215}
\author{Walter Gear \altaffilmark{4}}
\affil{Royal Observatory, Blackford Hill, Edinburgh EH9 3HJ , Scotland, UK}
\author{Harri Ter\"asranta, Esko Valtaoja}
\affil{Metsahovi Radio Research Station, Helsinki University of Technology,
Otakaari 5A , SF-02150 Espoo, Finland}
\author{Hugh D. Aller and Margo F. Aller}
\affil{Astronomy Department, University of Michigan, Dennison Building,
Ann Arbor, MI 48109}
\altaffiltext{1}{NAS/NRC Resident Research Associate, NASA/Goddard Space 
Flight Center}
\altaffiltext{2}{Current address: Infrared Processing and Analysis Center,
Jet Propulsion Laboratory and California Institute of Technology,
MS 100-22, Pasadena, CA 91125}
\altaffiltext{3}{Current address: Sterling Software, MS 210-2, NASA Ames Research Center,
Moffet Field, CA 94035-1000}
\altaffiltext{4}{Current address: Mullard Space Science Laboratory, Holmbury St. Mary,
Nr Dorking, Surrey RH5 6NT, United Kingdom}
\begin{abstract}
We present multiwaveband observations of a well-selected sample of 28 quasars
and two radio galaxies with flat radio spectra and strong millimeter-wave
emission (referred to here as FSRQ's).
The data are analyzed
to determine the radio to infrared and X-ray to $\gamma$-ray
properties of FSRQ's and the relationships between them.
Specifically, the synchrotron self-Compton (SSC) 
process is examined as a likely common radiation mechanism.  
For most sources, the broad band spectra are still incomplete,
especially in the far-infrared and ultraviolet range. Therefore
precise analysis, such as model-fitting of spectra is not usually
possible. To compensate partially for this, we have taken a statistical
approach, and examine the relationship between high and low
energy emission by using the data set for the entire sample. 

We use very long baseline interferometry (VLBI) at 8.4 and 22 GHz --- higher
frequencies than those of previous surveys --- in
conjunction with nearly simultaneous radio to submillimeter-wave 
observations to determine 
the parameters of the synchrotron spectrum and to examine the compact angular
structure of a subset of sources from our sample.
These parameters are used to predict the SSC X-ray flux densities.
Seven of thirty sources have predicted self-Compton X-ray flux densities 
well above the observed
flux densities obtained with the ROSAT satellite unless one assumes that
the radiating plasma
experiences bulk relativistic motion directed toward the
observer's line of sight.
Three of these seven sources are detected at $\gamma$-ray frequencies.
Model spectra show that the X-rays are consistent
with the first order SSC process, with the simultaneous multiwaveband
spectrum of the quasar 0836+710 obtained in 1992 March being
very well fit by SSC emission from a uniform, relativistically moving source.
The $\gamma$-rays are not produced via second order self-Compton
scattering, but rather by either first order self-Compton
scattering or some other process.

A comparison of the ROSAT X-ray flux densities and those obtained earlier
with the {\it Einstein Observatory} show that several FSRQ's are
X-ray variables on timescales of about a decade. Several sources 
that were observed more than once with ROSAT also show variability on
timescales of 1--2 yr, with
the X-ray variability in these cases often associated with 
millimeter-wave variability and lower VLBI core-to-jet flux ratios. 
Detections at $\gamma$-ray energies also appear to be related to increases
in the radio to millimeter-wave flux densities. 

Statistical analysis shows that the millimeter-wave and X-ray
luminosities for the sample are strongly correlated, with
a linear regression 
slope $\sim$ 0.6. The peak in the distribution of X-ray
to millimeter spectral indices also indicates a strong connection between
millimeter-wave and X-ray emission. Particularly interesting is a
correlation between X-ray to millimeter spectral index and fraction of
flux density contained in the VLBI core. This tendency toward higher
X-ray fluxes from sources with stronger jet emission implies that the
knots in the jet are prominent sources of X-rays.
\end{abstract}
\section{Introduction}
It is clear that in quasars there is a strong relationship between the
nonthermal emission at low frequencies and that at high frequencies. 
Quasars are generally strong X-ray emitters (e.g., \cite{tan79};
\cite{kuw80}), and
a number have been found to be strong, hard ($> 100$ MeV) $\gamma$-ray sources
(\cite{fic94}; \cite{tho95}; \cite{tho96}). Quasars with flat radio spectra --- which indicate compact,
milliarcsecond-scale structure --- and strong millimeter-wave emission (here
abbreviated as FSRQ's) have been demonstrated to be
more luminous in soft X-rays (for a given optical luminosity)
than either their radio-quiet counterparts
or radio sources with the same radio luminosity but
steeper radio spectral index (\cite{owe81}; \cite{zam81}; \cite{led85} ; \cite{kem86}; \cite{wor87}).
\cite{bro87} have also shown that for a given 
radio luminosity in extended (arcseconds or greater)
structure, the X-ray luminosity tends to be
more than an order of magnitude larger for sources with higher compact
radio luminosity. Furthermore, except for the Large Magellanic Cloud, bright 
extragalactic hard $\gamma$-ray sources
are all FSRQ's (or BL Lac objects, which have similar radio--infrared 
spectra). 

These strong correlations between radio and high-energy emission are
readily explained if the nonthermal emission at all wavebands is produced via
synchrotron self-Compton (SSC) scattering, as has been
suggested by many authors (e.g., \cite{jon74a}). This is,
however, not the only possibility, since other radiative processes are
likely to operate in an environment that produces the highly relativistic
electrons and strong magnetic field required for SSC emission (eg., \cite{der93}).

If the SSC process dominates the nonthermal emission, one would expect a
correlation between X-ray and $\gamma$-ray fluxes and
VLBI ``core'' fluxes.
However, a statistical study of the (non-simultaneous)
VLBI core emission and X-ray emission from compact radio loud quasars and
active galaxies (\cite{blo91}) did not conclusively demonstrate a strong correlation. Though Zhou {\it et al.} (\markcite{zho97}1997) do find a strong relationship between
high energy $\gamma$-ray fluxes and VLBI fluxes.
Nevertheless, detailed studies of individual sources, both with VLBI and
X-ray data, show that the SSC process is a plausible explanation for
the X-ray emission (\cite{mar88}; \cite{unw94}; \cite{unw97}; \cite{eck86}; \cite{eck87}).
The results of these investigations have motivated us to conduct a 
more detailed study of FSRQ's and the relationship between the
radio--infrared and the high energy radiation. The present study
not only explores the statistical relationships between
high and low energy emission, but also compares theoretical SSC spectra
to multiwaveband data from a set
of contemporaneous measurements. Observations over a large range in
frequency allow for more accurate determinations of spectral index,
spectral peak frequency and flux density, and consequently a more accurate
prediction of the Compton flux densities.

This study, which includes high-frequency (8.4 and 22 GHz) VLBI imaging, as
well as radio through $\gamma$-ray (not all wavebands for all sources) flux
density measurements,
constitutes the most comprehensive set to date of multiwaveband observations of
bright FSRQ's. Below we summarize the observations and
data analysis. We present our own
VLBI, X-ray, submillimeter-wave (submm), and near-infrared (IR)
data. A more extensive
presentation of previous radio--IR observations used in this work
is given by \cite{blo94b}.
We follow with explanations of the procedure for spectral
deconvolution and determination of spectral peak frequency and
peak flux density,  
SSC calculations for specific sources, and comparisons of model spectra
against the data. In the final sections we present a statistical
analysis of the sample and a general discussion of the
results. 
\section{Observations and Data Analysis}
\subsection{Source Selection}
The sample is taken from an essentially complete list of radio
sources stronger than 1 Jy at 5 GHz, with spectral indices between 1.5 and 4.9
GHz flatter than $\alpha = 0.5$ ($F_\nu \propto \nu^{-\alpha}$) 
(\cite{ste88}). Quasars and
galaxies with 90 GHz
flux densities exceeding 0.5 Jy, declinations $ \geq 0^\circ$,
and measured redshifts exceeding 0.03 were
selected from Table III of \cite{ste88} to generate a preliminary sample. This
resulted in a sample that had a preponderance of sources with redshifts
between 0.4 and 2.25. In order
to pare the sample down to a more reasonable size, within the
redshift range of 0.4 to 2.25 only sources with declinations $\geq 30^\circ$ were included in the final sample. BL Lac objects, which form a class distinct
from quasars with respect to their soft X-ray emission (\cite{wor90}) 
are excluded, as are optically unidentified objects. The final
sample contains 30 sources --- 28 quasars and two radio galaxies ---
and is well distributed in redshift. (One source that would have been in the
final sample, 3C 216, was dropped from the study due to its proximity to the Sun  
during the time of the submillimeter observations.)
Because of telescope time allocation limitations, it was not possible to observe 
the entire
sample at all wavelengths. Approximately one half of this sample was observed
from radio to X-ray wavelengths. This subsample was essentially determined
by which sources were observed (in priority order) with ROSAT prior to
the shutdown of the PSPC detector. This might then
have caused a selection of objects that were {\it a priori} deemed to be
more interesting than others. Both radio galaxies, 3C 111 and 3C 120, are included
in the subsample.
The list of sources, along with a summary of the observations at all
wavebands, is presented in Table 1.
\subsection{VLBI Data}
A subsample of sources has been observed with VLBI at
frequencies of 8.4 and 22 GHz. For this work,
seven observing sessions were conducted with either the Global VLBI Network or
the Very Long Baseline Array \footnote[1]{The VLBA is an instrument of the
National Radio Astronomy Observatory, a facility of the National Science
Foundation operated under cooperative agreement by Associated Universities, Inc.}(VLBA; after 1994). 
The stations of each array (defined by the epoch of the observations)
are listed in Table 2.
Sessions 1-6 used the Mark II VLBI recording system with a 
1.8 MHz bandwidth, while sessions 7-9 used the VLBA recording system
with 16 and 32 MHz bandwidth respectively.
Due to the large sample size, we 
observed in ``snapshot'' mode with each source typically 
observed for 4 to 5 scans of one-half hour each during the allotted observing 
time. 
Typically three to six sources were observed during each VLBI session.
These scan times were chosen to optimize
{\it u-v} coverage. The video tapes
from the individual antennas were correlated using the JPL/Caltech Block II
correlator for sessions 1-6 and the VLBA correlator for sessions 7-9.
All post-processing was performed at
Boston University using the NRAO AIPS, the 
CalTech VLB software package, and the difference mapping software Difmap. The 
phase
time and frequency derivatives (fringe delay and fringe rate) were
determined using
the global fringe fitting algorithm FRING in AIPS (\cite{sch83} ).
The amplitudes were calibrated in the usual manner as described by 
Cohen {\it et al} (\markcite{coh75}1975). Inconsistencies in the calibration 
of flux densities were corrected by using
{\it u-v} crossing points and then finding the best-fit flux density using a
${\chi}^2$ test.
A hybrid mapping procedure was used to obtain
images of the sources (see, e.g., \cite{pea84}). 
Hogbom's CLEAN algorithm was used to deconvolve the dirty beam from
the dirty image  to determine the true source brightness distribution.
A Gaussian restoring beam  with dimensions equal to those of the
central portion of the dirty beam was convolved with the
CLEAN components to produce a final map.

Table 3 summarizes the parameters of the final image for each source
at each frequency: elliptical Gaussian
restoring beam ($a_{\rm beam}$, $b_{\rm beam}$), the position angle of the 
beam 
($\theta_{\rm beam}$, as measured north through east), the peak flux density
per unit beam area, and the contour levels.
Table 4 summarizes the Gaussian model parameters used in 
conjunction with the hybrid mapping procedures, as described above.
These are from the model fits to the self-calibrated data. As with the
Pearson \& Readhead (\markcite{pea88}1988) survey, the limited data only 
weakly constrain some
model components. In Tables 3 and 4, $r$ is the angular distance from the map 
center,
$\theta_{\rm pos}$ is the angular position of the component relative to the
(arbitrarily placed) origin (measured from north through
east), $a$ and $b$ are the FWHM major and minor axes of the assumed elliptical
Gaussian brightness distribution of the component, and 
$\phi$ is the position angle of the major axis. The uncertainties in 
the flux densities given are the estimated calibration errors (as opposed
to the uncertainty in the model, which is usually larger). 
Though we have not, in general, estimated the uncertainties for
every parameter of each component, we have performed an uncertainty
analysis on the angular sizes (characterized as major and minor
axes of an elliptical  Gaussian here) of some sources. This will
be helpful later (\S 3.2) in determining the uncertainty of the predicted
Compton flux, which is highly sensitive to the angular size value.
In determining the magnitude of the 1-$\sigma$ uncertainities, we
use the technique of Biretta {\it et al} (\markcite{bir86} 1986)
Due to the large range in the data quality of our snaphot images,
there is also a fairly large range in the uncertainties of derived
angular sizes. For the 4 objects observed with the VLBA and also later used 
for the Compton calculations, the uncertainties are 13/
for 0945+408, 0955+476, 1611+343, and 2201+315 respectively.; however, for 
sources observed earlier
with the Global VLBI Network, the typical model uncertainties
are several times larger.    
Figure 1 displays the VLBI images at 8.4 and 22.2 GHz, in order of increasing
right ascension for sources mapped with Global VLBI. Figure 2 displays
the maps from the VLBA.
Note the differences in angular scale, especially between most of the 22 GHz
and 8.4 GHz maps. 
\subsection{Submillimeter and Infrared Data}
A detailed description of the data analysis at submillimeter and infrared
wavebands is given in Bloom {\it et al} (\markcite{blo94b}1994). In Tables 5 
and 6 we present data more
recently acquired with the JCMT and UKIRT. The last column in Table 5 is
a measure of variability amplitude, as determined by the parameter
${F_{\rm max}-F_{\rm min} }\over {F_{\rm max}+F{\rm min}}$. The two measured
flux densities for this calculation, extracted from the data presented
here or in Bloom {\it et al} (\markcite{blo94b}1994), are from the
two observations that are closest in time.
This measurement of the variability amplitude is considered to be
significant if there is at least a 2-$\sigma$ separation between the error
bars of the highest and lowest point. 
We have also examined the possibility of variability in the older data
(between 1991 and early 1993). The sources found to be variable are discussed
in \S 2.6.
\subsection{X-ray Data}
The Position Sensitive Proportional Counter (PSPC) was used in conjunction with
the Woltjer Type I nested mirror system onboard the Roentgensattelit (ROSAT)
to obtain fluxes over the energy range 0.1--2.5 keV. 
After initial processing at Goddard Space Flight Center (GSFC), the X-ray data 
were extracted from tape using the Post Reduction Off-line Software
(PROS) routines operating under the NOAO IRAF data analysis package. 
Before conducting a spectral analysis, the background counts were determined 
for each
source by calculating the number of photons within concentric annuli,
centered on the source. The precise sizes of the annuli (typically $2'$
inner and $3'$ outer) were determined
for each source by displaying the photons/pixel with SAOIMAGE (see \cite{wil92}). 
All of the objects in this sample are unresolved with the PSPC, which has a
spatial resolution of about $15''$. Therefore, spatial analysis
of the photon counts (i.e., mapping the source) is not possible.

The spectral analysis was conducted using the standard routines within PROS.
This consisted of modeling the data in terms of a power-law spectrum over the
energy range 0.1--2.5 keV, with
absorption at low energies due to intervening gas, parameterized as the 
column density of hydrogen atoms under the
assumption of cosmic abundances.
The X-ray photon spectrum is represented by the following equation:
\begin{equation}
\eqnum{2.4.1}
{dN \over {dE}}=K{E^{-(\alpha+1)}}e^{-N_H \sigma (E)}\qquad  {\rm photons \, 
s^{-1} cm^{-2} keV^{-1}}.
\end{equation}
The predicted spectrum 
(photon counts per energy
bin) using this model was then compared to the observed spectrum using
a ${\chi}^2$ test to determine ``goodness-of-fit.'' The model parameters
that minimize ${\chi}^2$ were adopted to calculate the total flux over a
specified energy band. The default values for the spectral response matrix
and effective area of the detector for routines in PROS 2.0 were used
to determine the source spectrum from the raw photon spectrum.
 Some care has to be taken, as ${\chi}^2$ can be a
minimum for parameter values that are physically unrealistic, 
such as an extremely high
column density, or a photon index that is much too steep. Uncertainties
in model parameters and derived flux values were determined by 
increasing the value of the parameter by the amount that increases 
$\chi^2$ to the desired confidence level. Alternatively, in some cases,
 the normalization
parameter, $K$, of equation (2.4.1) was plotted against spectral index, and 
the 
uncertainty
in this parameter was taken to be the range of the 68\%  confidence contour. 
The flux density was then calculated by using equation (2.4.1) corrected for
absorption, and then converting to Janskys.

Table 7 summarizes the ROSAT X-ray observations. Column (5) is the logarithm
of the Galactic hydrogen column density, ${N_H}_{\rm gal}$. Unless otherwise 
noted, this is the
value from the Stark {\it et al.} (\markcite{sta92}1992) 21-cm emission line observations.
However, for two sources,
3C 111 and NRAO 140, we have also added the column density (in atoms
cm$^{-2}$) of molecular hydrogen
as inferred from observations of Galactic CO (\cite{ban91}).
Column (6) gives the  
best fit value of $N_H$ as determined from the $\chi^2$ minimization routine. 
Column (7) is the
best-fit spectral index; if no uncertainties are given, then
it is an assumed value.
Column (8) is the 1 keV flux density. Columns (9) and (10) give the
$\chi^2$ and degrees of freedom (D.O.F., one less than the number of parameters varied subtracted from the total number of energy channels), respectively.
When $N_H$ is determined from a best fit, then the uncertainites quoted
in the spectral index and flux densities correspond to $\chi_{min}^2$ +3.53
(68\% confidence for three interesting parameters). If
$N_H$ is assumed to be the Galactic value, then the subsequent uncertainties
correspond to $\chi_{min}^2+2.3$
(68\% confidence for two interesting parameters).
If both $\alpha$ and $N_H$ are assumed then the uncertainties are
for $\chi_{min}^2$ +1.0 (68\% confidence for one interesting parameter).
 Generally, the fit is
considered reasonable if $\chi^2 \lesssim$ D.O.F. Note that by this criterion 
two fits are not
particularly good
(see Table 7). The fit for the first observation of 1150+812 is poor probably
because the fit was performed on the combined counts from
two scans that were a month apart, each having a low number of individual
counts. Also, the actual $N_H$ value could be smaller or larger than the
assumed Galactic value (see below for a discussion of this point).
It is also possible that a single power-law model
is not adequate for describing this source.
The first observation of 1611+343 is poorly fit. Probably only the last
reason is applicable here, since there are enough counts to determine the 
spectrum rather well, and the best-fit $N_H$ is fairly close to the Galactic 
value. 

In many cases the best-fit values for $N_H$ are either considerably smaller or
larger than the measured Galactic $N_H$. There are several likely explanations
for this. If the best-fit value is larger, as is the case for 3C 120 and
0736+017,
then there could be significant photoelectric absorption within the source. 
There could also be significant amounts of as yet unobserved Galactic 
molecular gas in the 
direction of the FSRQ. Also possible is a spectral flattening at low X-ray
energies. A closer look at the photon spectrum of 3C 120 shows that the
peak is poorly fit by the model, indicating that a single power-law is
probably not the best model for this source.
It is also possible that the Galactic column density of hydrogen is higher
(or lower) in the immediate vicinity of the FSRQ as compared to the average value over
the much larger regions measured with the $2^\circ$ FWHM beam of the 
Stark {\it et al.}(\markcite{sta92}1992) survey. This seems to be the case 
for 3C 345, for which
$N_H$ measured with the smaller beam in the Elvis {\it et al.} 
(\markcite{elv89}1989) study is 
$7.5 \times 10^{19}$, which is certainly closer to the best-fit value of 
$N_H$ for the 1990 observation than is the value
${N_H}_{\rm gal} = 3.6\times 10^{20} {\rm cm^{-2}}$ derived from 
Stark {\it et al} (\markcite{sta92}1992).
These possibilities have been discussed in more detail by Wilkes \&
Elvis (\markcite{wil87}1987)
and Worrall \& Wilkes (\markcite{wor90}1990), among others. However, the 
higher resolution H{\thinspace}I
observations of Elvis {\it et al.} (\markcite{elv89}1989) show
that ${N_H}_{\rm gal}$ is smaller than derived from 
Stark {\it et al} (\markcite{sta92}1992) 
for 0736+017 and only 2\% larger for 3C120. It is still possible that 
${N_H}_{\rm gal}$ in the immediate direction of the quasar was larger at the 
time of the X-ray observations, since temporal variability of ${N_H}_{\rm gal}$
cannot be excluded (see, e.g., \cite{fic94}).
For 0736+017, the inferred column density from CO observations
(\cite{lis93} ) is 5.6 ($\pm 0.8) \times 10^{20} ~{\rm cm^{-2}}$, 
which can 
account for the difference. This also results in a much steeper derived
value of the spectral index, which is more in agreement with other
low-redshift active galaxies and quasars (see discussion below). 
If the best-fit value of $N_H$ is smaller than the Galactic value,
as is the case for 0133+476, 3C 111,
1633+382, and 3C 345, then it is possible that these sources are not 
adequately described by a single power-law. The most likely possibility
is a soft X-ray excess, perhaps from 
thermal black body X-ray emission (from an accretion disk) with a steep 
slope. 

An analysis for some sources in this sample has appeared in other
summaries of X-ray emission from blazars (see \cite{com97}; \cite{sam97}) . 
Most of the parameter determinations (such as spectral index and
column density of hydrogen) are very similar. Our flux density values
do occasionally differ, mainly because we base our flux values on best
fit parameters, whereas Sambruna (\markcite{sam97}1997) use fixed ${N_H}_{\rm gal}$ to calculate
flux densities. In addition, their source/background counts determination
differs from ours slightly. The first observation of 1611+343 is the only
case for which the flux densities differ by more than 5\% (we present a value
which is 50\% lower).
\subsection{Gamma-Ray Data}
Several sources in our sample were detected with the EGRET instrument
on the {\it Compton Gamma-Ray Observatory}. The fluxes for EGRET
detected objects can be found in Thompson {\it et al.} (\markcite{tho95}1995) 
and upper limits for
some sources in ths sample can be found in Fichtel {\it et al.}
(\markcite{fic94}1994). The dates listed in Table 1 are for the EGRET
Viewing Periods, covering the area of the source, which were closest in
time to the other multi-wavelength observations. 
Calibration of the instrument
and determination of fluxes are discussed in Hartman {\it et al.} (\markcite{har92}1992),
Mattox {\it et al.} (\markcite{mat93}1993), and 
Mattox {\it et al.} (\markcite{mat96}1996), respectively.
\subsection{Summary of Source Properties}
Due to incomplete time coverage, it is not possible to analyze thoroughly
the variability properties of each source. However, we can discuss some
general results. Of the 17 sources with repeated millimeter/submillimeter
wave observations reported here and in Bloom {\it et al.} 
(\markcite{blo94b}1994),
ten (0133+476, NRAO 140, 3C 111, 0642+449, 0736+017,
0836+710, 3C 345, 2005+403, 2037+511, and 2201+315)
are significantly variable at one or more
wavelengths. Six of these (0133+476, NRAO 140, 0736+017, 0836+710,
3C 345, and 2201+315) are significantly variable at all measured
wavelengths. The timescales are roughly
7 to 14 months; no source has a variability amplitude parameter that
exceeds 0.5. 

With the X-ray observations presented earlier, combined with observations
by previous spacecraft missions (HEAO A-1, Einstein, EXOSAT, and Ginga),
we have determined the crude X-ray variability characteristics of the sources
in the sample, using the same criterion for significance as for the
millimeter observations. We note that in some cases it is difficult
to establish variability, especially if the observations being compared
were made with different instruments or very different spectral fit
parameters (as mentioned above). 
The sources appear to fall into several general
categories (\cite{mal94};\cite{blo94a}  for 
references to the published X-ray data for each source). Of the 24
sources in our sample with ROSAT X-ray observations,
seven (0133+476, 0552+398, 0820+560, 0917+499, 0954+556, 0955+476,
and 1638+398) have indeterminate X-ray variability due to insufficient
observations. For an additional seven sources
(0016+731, 0642+449, 0736+017, 0804+499, 1150+812,
2136+141, and 2201+315) variability was not detected, though some
of these are low amplitude
variables at mm wavelengths (e.g., 0736+017; see \cite{tor94}).
Two sources, 0945+408 and 1633+382, are observed to be significantly
variable in X-rays (and, for 1633+382, $\gamma$-rays; see \cite{tho95}) only, whereas three additional sources (NRAO 140, 0836+710
and 4C 39.25) show possible X-ray variability that is temporally correlated
with radio or mm-wave variability. With each of these three sources,
the mm flux varied by the same factor as the X-ray flux
(see \cite{mar88} for NRAO 140; this work and Bloom {\it et al.}
(\markcite{blo94b}1994) for
0836+710 and 4C 39.25). These last five sources mentioned
were all variable with constant X-ray spectral index, which is typical of
nonthermal processes. 
Five additional sources
(0212+735, 3C 111, 3C 120, 1611+343, and 3C 345) had complex variability 
properties, such as variation in spectral index
as well as intensity. All of the sources with $\lesssim$ 80 \% of the VLBI
flux density in the core at 8.4 GHz are X-ray variable, falling in one of
the last three categories. The possible relationship between X-ray and
milliarcsecond-scale structure is discussed in \S 4.2 . 
\section{Spectral Analysis}
\subsection{Spectral Deconvolution}
In this section, the radio to submillimeter and infrared spectra
of Bloom {\it et al.} (\markcite{blo94b}1994) are used 
to determine the synchrotron spectral turnover frequencies for each source in
the sample with sufficient data. In addition, for sources observed with
both ROSAT and VLBI, X-ray 
measurements are compared with the theoretical 
spectra of synchrotron self-Compton (SSC) models generated
by the techniques described below. For sources with published EGRET $\gamma$-ray
data, those data are also included for further comparison with the theoretical 
spectra.
The total flux density  spectra presented in Bloom {\it et al.} 
(\markcite{blo94b}1994) are dissected into the
spectra of individual VLBI scale components using the total flux density data 
and flux densities from the best-fit models
to the 8.4 and 22 GHz VLBI visibilities. The emphasis of the spectral 
decomposition is
on finding the spectral turnover frequency and flux density of the bright 
``VLBI core'' component usually located at one end of a source.
Due to the limited data and the weakness of the secondary 
components we do not attempt to fit the spectra of any non-core ``knots''
(except for the peculiar source 4C 39.25,
which was analyzed by \cite{zha94b} as part of a separate study).

In Figure 3 the spectral dissection for a representative
source is shown.
The turnover frequency $\nu_m$ is determined from the data by first calculating
the frequency $\nu_n$ at which 
the optically thin and optically thick spectra (of slope $\alpha_{\rm mm}$
and $\alpha_{\rm thick}$, respectively) intersect
(\cite{jon74b} ; \cite{mar77}). In some cases this 
is not straightforward, since the VLBI spectra are complex (e.g., 3C 120 and
1633+382). For such sources, there is significant 
blending of components at 8.4 GHz that are separated in the higher resolution
22 GHz maps. There are also additional components not visible in the 22 GHz images
because of over-resolution and/or steep spectra.
There are insufficient data to separate these components properly, thus only
the 22 GHz core flux is used with the total flux density data to determine
the spectral turnover. 
In sources with no VLBI data, best estimates are made using the total flux
density data. We also note that in some cases, the VLA and/or VLBI
data points exceed the values for the single-dish measurements (Figure 3). 
Since
it is obviously not possible for the small scale components to have greater
flux density than the entire source seen by a single dish,
these differences in flux are attributable to differences in calibration.
Since all of the sources have falling millimeter and
submillimeter spectra, there is a firm upper limit of 100-150 GHz to the
turnover frequency. All sources in the sample have optically thick
spectral indices below the turnover that are much flatter than the value
of 2.5 expected from a homogeneous source. These flatter indices are 
indicative of sources with decreasing radial gradients in magnetic field $B$
and electron energy distribution normalization factor $N_0$.
Using the expressions of Marscher (\markcite{mar77}1977), who considered such 
cases, the actual synchrotron turnover frequency $\nu_m$ is related to the 
intersection
value $\nu_n$ by a constant factor, which in most cases is close to unity.
The quantity $F_m$ is the turnover flux density determined by extrapolating the
optically thin spectrum down to
frequency $\nu_m$. The values of $\nu_n$, $\nu_m$, $F_m$, $\alpha_{mm}$, and 
$\alpha_{\rm thick}$ for each source are listed in Table 8. 

For sources observed with VLBI, 
the angular sizes are derived from those obtained from
the models that best fit the visibilities as described in \S 2.2.
Though elliptical Gaussians are used to represent the brightness
distribution of each source component, for the SSC calculations
the visibilities can equivalently be represented by a uniform sphere 
with angular diameter $\theta_{us}= 1.8\sqrt{\theta_a \theta_b}$
(\cite{mar87}). The equivalent size for a spherically symmetric, non-uniform
source is 
$(1.8 \, k_{\theta})^{-1}\theta_{us}$, where $k_{\theta}$ is a
parameter in the range 1--2 (\cite{mar77}). The value $\theta_{us}$ is used
in the subsequent modeling calculations. This angular size value is
only strictly valid if the core is known to be optically thin. In most
cases, our spectra reveal the core to be optically thick; however, we
use these optically thin values for consistency with the following
Compton calculations (which assume the source is optically thin). 
\subsection{Synchrotron self-Compton Calculations}
Once $F_m$ and $\nu_m$ are determined, these values as
well as millimeter spectral indices and angular sizes are used with
the following equation (multiplied by a constant factor; see \cite{mar87} ):
\begin{equation}
\eqnum{3.2.1}
{F_\nu}^{1C} \propto {\theta}^{-2(2\alpha_{mm}+3)}{F_m}^{2(\alpha_{mm}+2)}{\nu_m}
^{-3\alpha_{mm}+5}{E_{keV}}^{-\alpha_{mm}}ln(\nu_2/\nu_m)\Biggl({{1+z}\over{\delta}}
\Biggr)^{2(\alpha_{mm}+2)}. 
\end{equation}
Here, $\nu_2$ is the upper cutoff frequency to the synchrotron spectrum
and $\delta$ is the Doppler factor corresponding to bulk relativistic
motion. We use eq. (3.2.1) to predict the X-ray flux density at
1 keV under the initial assumption that there is no relativistic beaming
($\delta = 1$). The angular sizes used in this
equation should be measured at the turnover
frequency. In general this is not the case, since VLBI observations
were only undertaken at 8.4 and 22 GHz, below $\nu_m$ for most sources.
In the case of a turnover frequency $\nu_m > 22$ GHz, the measured angular
size is expected to follow the inequality $\theta(\nu) > \theta(\nu_M)$,
and the Compton flux is underestimated. If the
source is transparent at 22 GHz, then the measured angular size is 
acceptable.  For sources with IR data, the
cutoff frequency $\nu_2$ can be estimated, and is typically
$\sim 10^{14}$ Hz. Otherwise, this frequency is assumed to be $10^{14}$ Hz.

In a number of
cases the X-ray flux densities predicted in this way are much greater than
those
measured. If relativistic beaming is taken into account, a minimum
Doppler factor can be derived from the measured and
predicted X-ray flux densities:
\begin{equation}
\eqnum{3.2.2}
{\delta_{min}}={\Biggl( {F_{\nu x,obs} \over F_{\nu x, pred}}\Biggr)}^
{-{{1 }\over {2(\alpha_{mm}+2)}}}.  
\end{equation}
The best estimate of the Doppler factor is then used with the initial
estimate of the X-ray flux density to generate model first and second
order SSC spectra. Model spectra for selected FRSQ's with minimum
Doppler factors obtained from the SSC calculations
are shown in Figures 4 (0133+476) and 5 (0836+710).
The parameters used to generate the models
are listed in Table 9. The SSC calculations show that seven of sixteen sources
observed with both VLBI and ROSAT have predicted X-ray flux densities at 1 keV 
greater than the observed values. The resolution of this discrepancy
requires relativistic beaming of the radiation. 

To evaluate more clearly what these predictions mean, we can roughly 
estimate the uncertainty in the SSC predicted flux and the Doppler
factor using standard propagation of errors techniques
(i.e., \cite{bev69}). Though
this strictly only holds for parameters which have normal distributions
and symmetric errors, a meaningful result for the uncertainty can be
evaluated if we treat the terms of eqs. 3.2.1 and  3.2.2 logarithmically:
\begin{equation}
\eqnum{3.2.3}
{\sigma_{tot}^2}={a^2}\sigma_{log \theta}^2 +{b^2}\sigma_{log F}^2 +
{c^2}\sigma_{log \nu}^2 + {...}.
\end{equation}
Here, $\sigma_{tot}$ refers to the total logarithmic uncertainty in
the predicted Compton flux. $a$, $b$, $c$ refer to the powers involving
spectral index which appear in eq. (3.2.1). The corresponding subscripts
refer to angular size, maximum synchrotron flux density and the spectral
turnover frequency. There are then other smaller or unquantifiable
terms which we have not written down (but we discuss them below). 
Following this technique, and assuming approximately 50 \% errors in
the angular size, and 40\% errors in the turnover flux and frequencies,
and a spectral index of approximately 0.7, the uncertainty in 
log $F_{\nu x, pred}$ is about 2. Reallistically, since the
spectral index itself can be uncertain by a few percent, and since there
are very likely to be correlated error terms which we haven't included, this
number could be  closer to 3. Though correlated error terms could also
reduce the overall uncertainty, in the absence of more information we
feel that it is wise to adopt the larger number. This number
corresponds to a ``worst case scenario''. As discussed earlier in
\S 2.2, some of the VLBI data have angular sizes with uncertainty
as small as 13\%. The resulting uncertainty would be smaller by more
than an order of magnitude in those cases ($\sim 50$). However, in general,
over-estimations of the
Compton flux, are only significant if this is an over-evaluation
by more than 3 orders of magnitude. By this standard, all of the
over-estimates in Table 9, except for 0133+476, are marginal, however many of 
the underestimates
are not. Using similar analysis on $\delta_{min}$, and realizing the
$F_{\nu x, pred}$ is the main contribution to the uncertainty, we
calculate that the uncertainty in the log of the minimum Doppler
factor is about 0.3. Thus, we conclude that the uncertainites 
are too large, generally, to determine whether Compton scattering
and Doppler beaming are necessary to reconcile predictions with the
X-ray data.

In some cases the X-ray spectral slopes predicted by the SSC mechanism
disagree with the slopes determined from fits to the ROSAT data. This could 
be due to non-SSC contributions (e.g., from a
``soft X-ray excess'' component) or simply a poorly derived X-ray spectrum
arising from the limited energy range to which the ROSAT PSPC is
sensitive. Nonuniform models (e.g., shock waves in which the electrons
suffer radiative losses; \cite{mar85}) can produce
self-Compton X-ray spectra that are flatter than the optically thin
synchrotron spectra.

For 0836+710, the $\gamma$-ray spectrum of 1992 March is roughly in
agreement with the predicted composite first and second order
multiwaveband spectrum (Figure 5). These results are similar, in general,
but not in detail, with the modeling results for 0836+710 presented by
Comastri {\it et al.} (\markcite{com97}1997).
The model predicts that the source
should have become faint at $\gamma$-ray energies by 1992 Oct-Nov,
and indeed no detection was reported when the source was next
observed in early 1993 (\cite{tho95}). This illustrates
the value of contemporaneous multiwaveband observations in testing
models: source variability invalidates tests based on non-contemporaneous
measurements.

In addition to modeling the spectra, we use observed synchrotron parameters,
such as $F_m$, $\nu_m$, $\alpha$, $\nu_2$, and $\theta$, in Table 10
to determine
physical parameters such as $N_0$ and $B$. 
These results show that the magnetic field $B$ ranges from
$1 \times 10^{-2}$ to 140 $({{\delta_{min}}/\delta})$ G  and $N_0$ from 
$9 \times 10^{-13}$ to $2 \times 10^{-3} {(\delta/\delta_{min})}^{2(\alpha+2)}
$ (cgs units).

For sources previously observed to exhibit superluminal motion,
the range of possible values for the Doppler factor derived from the apparent
transverse velocities $v_{\rm app}$ can be compared to 
those derived using eq. (4.2.2). Known superluminal sources with measured
X-ray fluxes lower than the predicted values tend to have  minimum
required values of $\delta$ that are compatable with the observed values
of $v_{\rm app}/c$,
as should normally be the case: 0836+710 ($v_{\rm app}/c = 15 h^{-1}$ [\cite{kri90}], $\delta > 3$),
4C 39.25 ($v_{\rm app}/c = 3.5 h^{-1}$, $\delta >5.2$
[\cite{mar91}], and 3C 345 ($v_{\rm app}/c = 8 h^{-1}$,
$\delta > 2.1$; see \cite{unw94}).  (Here we adopt
the values ${\rm q_0}$=0 and $H_0$ =100$h$ $\rm km ~s^{-1} ~Mpc^{-1}$.)
However, this need not be the case for all sources,
since the value of $\delta$ can be high while $v_{app}/c$ is low for
jets pointing essentially along the line of sight. In such cases,
the jet should not appear as extended as in the high $v_{\rm app}/c$        
sources. 
\section{Statistical Analysis}
We use correlation and linear regression
to test for and to model approximately the bivariate relationships
among the data. We employ these techniques with emphasis on those parameters
most likely to be 
related to X-ray emission and $\gamma$-ray emission. Since the SSC process
predicts a specific relationship between high energy flux densities, radio 
flux density, spectral turnover frequency, and angular size, these parameters are
studied most closely. In addition, other parameters such as the X-ray and millimeter-wave spectral indices, and  VLBI compactness, are examined.

All of the luminosities in this section are calculated assuming
${\rm q_0}$=0 and $H_0$ =100$h$ $\rm km ~s^{-1} ~Mpc^{-1}$,
with $h$ arbitrarily chosen to be 0.75.
\subsection{Survival Analysis--Correlation and Regression}
Due to the presence of upper and lower limits to some of the measured or
inferred quantities (data with limits are often referred 
to as ``censored'' data),
survival analysis techniques
are used to evaluate more accurately correlation
coefficients (with their associated probabilities) and regression slopes
(\cite{iso86}). In general, these techniques take
into account censored data
by assuming a form for the distribution of the censored values relative to
the measured ones.
 The ASURV Rev 1.1 
statistical software 
package is 
used to calculate correlation properties for censored data
(\cite{iso90}).  As pointed out by \cite{iso86}, spurious 
correlations (or spurious null correlations)
can occur if these limits are disregarded.

There are many correlation tests and linear regression methods.
In general, since they are both non-parametric, the Spearman rank correlation 
coefficient and Kendall's
$\tau$ are best suited  to test for correlation among two variables for which
the nature of the parent distributions is unknown. 
The  Spearman rank coefficients are provided in the summary of our results in
Table 11. 

The Buckley-James method (BJ) is used for determining regression coefficients.
This method is preferred over other methods, since it does not assume a 
Gaussian distribution
of the residuals about the regression line. In cases for which there are limits
present in the independent variable (e.g., the possible dependence of X-ray flux density on angular  size), only Schmitt's 
regression method (SB) can
be used. However, this method is not considered preferable for general use,
since it involves arbitrary binning of the data.
(For a more detailed discussion of the relative merits of these methods see 
\cite{iso86}).
When no limits are present, the regression results are nearly identical to 
the results from standard methods (e.g., \cite{bev69}).

Since several previous studies have shown a strong correlation between 
X-ray and radio luminosities for radio loud quasars 
(e.g., \cite{wor87}), we examine this relationship first.
We consider separately the correlation between X-ray luminosity and radio
core luminosity, X-ray and millimeter luminosity at 270 GHz, and X-ray and
radio luminosity at spectral turnover. The analagous relationships between
the flux densities as well as between  the flux density at each waveband and
redshift are studied, to ascertain whether the luminosity correlations are
intrinsic or spurious.

As can be seen in Figures 6--11 the X-ray and millimeter flux densities are
correlated, the X-ray and radio core flux densities are uncorrelated, and the X-ray and turnover flux densities are weakly correlated.
All of the luminosity-luminosity plots in these figures show moderate or
strong 
correlations. The X-ray/millimeter relationship may be stronger
than that of the X-ray/core because the core is optically thick.
Optically thick synchrotron emission only makes a very small 
contribution to the total soft radiation field which is upscattered
to X-ray energies (\cite{ban86}).  
A summary of the correlation and regression analysis is given in Table 11.
Regression results are given only if the probability that a correlation
is not present is less than 20\%.

If the correlation between the luminosities were due solely to the common 
distance factors used to derive them, then the regression slope would be 
close to
unity. However, the regression slope is less than one in all cases. 
Isobe {\it et al.}(\markcite{iso86}1986)
show through Monte Carlo simulations of data from flux limited samples
that if the underlying physical
relationship between the  luminosities is of the form ${L_2}=KL_1^\alpha$,
then luminosity diagrams will reflect this through the regression slope,
as long as upper limits are included in the analysis. The resulting
flux--flux plots show a correlation with considerably steeper slopes. 
Although the underlying
physical relationship between the luminosities in this sample is likely to 
be more complex than the one assumed by Isobe {\it et al.}
(\markcite{iso86}1986)
their results do indicate that, for some forms of luminosity functions, 
the data will show the intrinsic correlation in the luminosity-luminosity
plots even if  the sample has a distance-luminosity bias. 
To evaluate the possibility of spurious luminosity correlations due
to common redshift factors, we have used the partial correlation
methods of Padovani (\markcite{pad92}1992). This analysis calculates the 
correlation
between two variables, while effectively removing  the effects of other
known variables. We find that there is still a moderate
correlation between $L_x$ and both $L_{mm}$ and $L_c$
($\rho_s=0.28$ and 0.39 respectively) , after the
correlations of each luminosity with redshift have been negated.  
However, the partial correlation between $L_x$
and $L_m$ is weak, indicating that the apparent relationship between these
two luminosities found earlier is probably spurious. The results are not
strongly affected if the extremely low redshift objects 3C 111 and 3C 120
are removed from the sample (as is seen in Table 11). 

Figures 12--17 give correlation plots and histograms of various other interesting
observational parameters.
Owen {\it et al.} (\markcite{owe81}1981) used the
histogram of the effective millimeter to X-ray spectral index to show that
the X-ray fluxes are closely linked to the millimeter fluxes. 
Bloom \& Marscher (\markcite{blo91}1991) showed that this also holds in 
general for the
radio cores of quasars, even though the observations were often separated
by years. Figure 15 shows that this is also true for the present sample,
with millimeter and X-ray observations separated by 6 months at most.
Inspection of Fig. 12 and Tables 11 \& 12 show that there is a weak 
anti-correlation
between X-ray spectral index and both millimeter and infrared spectral
indices. This would tend to show that a harder component to the X-ray
emission is visible when the synchrtron spectrum 
is steep, and not contributing as much to the X-ray flux via the SSC 
mechanism. This finding is somewhat surprising, since we would 
expect that sources with less significant SSC emission would then
be dominated by the softer (steeper) spectrum of an accretion disk.
Though, since several of the sources with the steepest millimeter spectral 
indices are at high redshift, it could be that the softest X-rays from
an accretion disk are redshifted out of the ROSAT band.
The anti-correlations
of the spectral indices are
especially interesting considering that the fluxes in the millimeter and X-ray
wavebands (and the luminosities) are correlated. It would be hard to argue in 
favor of the SSC mechanism (at least in a simple form) if these correlations
are observed in the future for simultaneous data from a large set of objects.
However,
without detailed X-ray spectral analysis of individual sources
(which will have to await AXAF) it is difficult to analyze this result
much further.
There is also a strong correlation between $\alpha_{mmx}$ and VLBI core
fraction at 8.4 GHz, as seen in Figure 16.
Here, core fraction is defined to be the fraction
of total milliarcsecond-scale flux density (as determined by model fitting) that is
in the core component. Note that there is one outlying point, which corresponds
to 3C 120. Even with 3C 120 excluded the correlation is highly significant
and the regression relation remains similar to the earlier case.
The correlation weakens at 22 GHz (Fig. 17). Although this may
be due to the inferior calibration at high frequencies, it is also true
that the jet (as opposed to core) components tend to be weaker, and
the core stronger, at 22 GHz than at lower frequencies.
The correlation suggests that more X-rays are created for a given
mm-wave luminosity if jet components are more prominent. That sources with
more prominent jet components should be brighter at high energies
was also suggested by Wehrle \& Mattox (\markcite{weh94}1994) in their correlation between
VLBI structure and $\gamma$-ray detections, although their result was of
marginal statistical significance. It is certainly not expected if the
bulk of the high-energy emission comes from the quiescent jet, as in
the Ghisselini {\it et al.} (\markcite{ghi85}1985) model.

There are various explanations for this correlation.
If the bulk Lorentz factors for each source are similar, then the
sources with smaller angles between the jet axis and the line of sight will
have larger
Doppler beaming factors.  Such sources will be more severely foreshortened
as viewed by the observer, and thus the jet ``knot'' components will not
be as easily detected (because of blending with the core)
as for jets that are less highly beamed. This effect will
also tend to decrease the SSC X-ray flux for a given
millimeter flux (all other quantities, such as turnover frequency, being
roughly equal; see eq. [3.2.1]). 
Recent VLBI and X-ray observations of 3C 345 (\cite{unw97}) indicate
that there are states in which the X-ray emission emanates from
the jet and not the nucleus, as one might normally expect. At these
times the relative contribution of radio flux from the extended jet 
is much higher. However, we note that $\alpha_{mmx}$ was uncorrelated 
with the core fraction as this particular flare progressed in 3C 345.

Another possible explanation is that
the sources with less prominent jets have wider opening angles, so
knots expand rapidly, which would decrease their SSC emission.
If this is the case, such sources should be found to be more highly
variable (at least in the decay stages of flares) than those with
prominent jets.
It may also be that the sources with strong jets are in an enhanced
state (e.g., from shock compression), which would increase the ratio of
X-ray to millimeter flux as compared to the other sources.
However, for several individual sources (NRAO 140, 0836+710, and 4C 39.25),
this ratio seems to be preserved during flares. 
As mentioned in \S 2, long term X-ray variability appears to be related
to 8.4 GHz VLBI structure as well, though this information is probably
too limited.
In the Melia \& K\"onigl (\markcite{mel89}1989) model, the X-rays arise from inverse Compton scattering of
seed photons from an accretion disk by relativistic electrons deep in
the jet. Because of the geometry of the scattering, the X-ray emission
is lower along the jet axis than at a small angle (roughly the inverse
of the Lorentz factor of the jet); viewed from this angle,
the jet components would have maximum visibility.

There are no significant correlations between X-ray flux density and the other
parameters used to predict the X-ray flux. This is somewhat surprising since
one might expect strong correlations from relations such as eq. (3.2.1).
A lack of correlation between X-ray flux density and angular size can,
however, be understood since many of the angular sizes are actually upper 
limits,
and other angular sizes may represent the blending of two or more components.
Futhermore, the actual angular sizes at the turnover frequency are likely
to be smaller than those measured at 22 GHz (see \S 3). 
The weak correlation between X-ray flux density and turnover frequency
can be understood since many of the turnover
frequencies are determined with limited data and are approximate. 
This uncertainty would then also be exacerbated by an inaccurately
determined turnover flux density. 
Of course, the simplest explanation for all of these non-correlations is that the X-ray flux
is not SSC in origin, and thus none of the correlations would be strong. If
this is the case, the actual Doppler factors must exceed the minimum values
obtained from the SSC analysis (Table 9).
\subsection{Distributions}
The histograms of the observed properties of FSRQ's show which values
are most common among the objects. These reveal both selection effects
and the physical aspects of the objects, as is demonstrated below.
The distribution of turnover frequencies (Fig. 13)
shows a very broad peak at roughly 50 GHz, similar to that found by
Gear {\it et al.}(\markcite{gea94}1994).
The distribution cuts off well below 150 GHz at high frequencies.
%
%
If the flat spectra are caused by a jet with radial gradients in magnetic
field and electron density (\cite{mar77} discusses the spherical case;
see \cite{mar80}; \cite{kon81} for application to jets), then the angular 
size of the emission region, observed in the partially opaque portion of the 
spectrum, decreases inversely (the precise dependence is a function of the
gradients in B and N) with frequency up to the spectral turnover.
Thus, the
spectral turnover frequencies are indicators of the innermost radius of the 
source. 
That the spectral  turnovers occur at frequencies $\sim$ 50 GHz, rather than 
up to an order of magnitude higher (as has been observed for some BL Lac
objects and flaring quasars; see \cite{bro89}; \cite{gea94}) 
indicates that the smallest radius is
farther out from the central engine than is implied for these other 
sources. The relationship $\theta_m \propto  \theta_{22}(\nu/22 GHz)^{-k_2}$
is used
to approximate the size of the inner core for the sources in our sample, where
$\theta_{22}$ is the angular size measured at 22 GHz. If that size is an upper 
limit, a value equal to half of the upper limit is used. The exponent $k_2$
is a constant of order
unity that depends on the optically thick spectral index and 
is enumerated in Marscher {\it et al.}(\markcite{mar77}1977).
A histogram of this estimated  ``innermost radius'' is  provided in 
Figure 14. The peak in the distribution of turnover frequencies (see Fig.
13) is an indicator
of the peak in the histogram of the inner radius at $\sim 0.6$ parsecs.
Though, considering our previous discussions (\S 2.2; \S 3.1-3.2) regarding the uncertainty
of the angular size, assuming an optically thin/thick source, and possible
blending of components, we warn that this number should be seen as 
a rough estimate of where a physical change in the jet should take
place and not as a canonical number.
The physical meaning of this inner radius is discussed in
\S 5 below.

The distribution of millimeter spectral indices (see Table 12) has a sharp
peak at $~\sim$ 0.7, and is clearly skewed toward steeper spectral indices.
\section{Discussion }
We have shown that the high energy properties of FSRQ's
are related to their radio to infrared properties. In particular,
the mm--X-ray luminosity correlation is strong, as found in previous
studies. In addition, we find an
anti-correlation between X-ray flux and VLBI core dominance, which suggests
that the knots which comprise the observed radio jets in FSRQ's
contribute substantially to the X-ray emission.

That the distribution of spectral indices peaks near $\alpha_{mm} \approx 0.7$
indicates that most of the sources contain  a distribution of
relativistic electrons with a canonical 
$E^{-2.4}$ spectrum. However, several sources have
much steeper millimeter-wave spectral indices, as expected for radiative
energy losses, or emission from an inhomogenous jet
(\cite{mar80}). This point is further borne out by the infrared observations.
Of the fifteen sources for which we have obtained infrared data
(reported here in Table 6 and in \cite{blo94b}), six show
infrared spectral indices that are 0.5 to 1.0 steeper than the mm-wave spectral
indices, which is again consistent with the basic predictions for
radiative energy losses. However, seven sources have infrared spectral indices
that are flatter by similar amounts, probably because of another (thermal?)
emission component becoming prominent toward higher frequencies.

The combined peaks in the distributions of spectral turnover frequency
and angular size are shown above to indicate a
characterictic size of about 0.6 pc for the radio cores of the FSRQ's 
observed with VLBI. In an inhomogeneous jet model with radially decreasing
magnetic field and electron energy normalization, the synchrotron
turnover frequency is predicted to increase with decreasing radius 
(\cite{mar80}; \cite{ghi85}).
Hence, if jets extended down to some arbitrarily small radius, 
this would result in much higher observed turnover frequencies.
Thus, there is clearly some physical inner cutoff to the radio jet.
The exact nature of this is unknown.
There could be a change in geometry at this radius.
That is, the jet could make a transition from a conical geometry to one which
has a weaker dependence on radius, such as a paraboloid (e.g. \cite{mar80}). Magnetic field and 
electron density, and therefore the turnover frequency, would then depend 
less strongly on radius. 
This transition in geometry could be caused by a change in the external
pressure on the jet or the cooling of the protons to nonrelativistic
mean internal energies.
A change in Doppler factor along the jet, either from a change in jet speed or
orientation, could also have the effect of
creating a cutoff in the emission since the observed flux from 
the jet follows $F_\nu \propto \delta^{3+\alpha_r}$.
Millimeter-wave VLBI observations are important to study this transition
region.

The lower redshift objects appear to have distinct X-ray properties.
For example, both 3C 111 and 3C 120 have very steep X-ray spectra
($\alpha_x> 2$) and 
the X-rays could have a thermal origin (\cite{der93}; \cite{gra97}). This steep
soft X-ray spectral component would lie at frequencies below the ROSAT
PSPC response at the higher redshifts of the quasars in the sample.
Indeed, at higher X-ray energies both radio galaxies have flatter spectra, 
similar to quasars (\cite{pet84}). Leach {\it et al.} (\markcite{lea95}1995) 
have shown that the spectral 
slope of the low-redshift quasar
3C 273 is $-2.7$ in the low energy band (0.1--0.3 keV) of the ROSAT
PSPC and $-0.5$ in the hard band (1.5--2.4 keV).
Also, as discussed earlier,
0736+017, at a redshift of 0.19, has an X-ray spectral index $\gtrsim$2
if photoelectric absorption by Galactic molecular gas is taken into account.

Three sources, 0836+710, NRAO 140, and 4C 39.25, clearly show radio to 
millimeter variability that is directly proportional the X-ray variability. 
As discussed by Marscher (\markcite{mar88}1988), this would occur if the 
quantity
$N_0 x$ is nearly constant with time, with $N_0$ varying inversely as
the line of sight distance through the source $x$.
However, an increase in the Doppler factor would also have a similar
signature (see \cite{mar92} and below). It is further 
likely,
from the general considerations of Bloom \& Marscher ({\markcite{blo96}1996) 
and the SSC 
calculations in
\S 3, that the high states of radio to X-ray (and in some cases
$\gamma$-ray) emission are associated with higher high-energy cutoffs to the
electron distribution.
This increases the X-ray flux density by another factor of
$2 \ln \gamma_{max}$. This would predict that if the $\gamma$-rays
are from first order SSC, then
the X-ray flux density should increase by a slightly larger factor than the
synchrotron flux density. In cases for which the entire spectrum increases
by a fixed amount, the increase can be attributed to a change in the
Doppler factor.

Models that involve the scattering of external radiation from the relativistic
electrons in jets (e.g., Melia and K\"onigl \markcite{mel89}1989; 
Dermer \& Schlickeiser \markcite{ders93} 1993)
would also explain this linear relation 
between millimeter and X-ray flux if changes in $N_0$ were solely
responsible for the variations in the synchrotron and scattered fluxes.
This would be the case if density fluctuations occurred
while the magnetic field remained constant (see \cite{blo96}).

Of the six sources in our sample that were detected in high energy
$\gamma$-rays, two were in a higher state at mm or cm wavelengths
at the time of the detection.
0836+710 showed a factor of two increase in the millimeter and X-ray
flux densities when it was detected in $\gamma$-rays relative to earlier
times. Though 0804+499 was only
observed at millimeter wavelengths several months before and
after the $\gamma$-ray detection, cm-wave flux densities clearly
increased by a factor of 3 between August 1991 and December 1992. The 
other sources (0917+449, 0954+556, 1611+343, 1633+382) were not well observed 
during the time of the EGRET observations.

The physical relationships among the various parameters are often
complicated and hidden by selection effects. We have used statistical tests
to attempt to reveal underlying relationships. We find that the X-ray luminosity is 
correlated with the radio core and millimeter-wave
luminosities. However, the strong correlation is probably due in part
to luminosity--redshift bias.

There are many aspects of this work that should be continued to gain
a better understanding of the relationship between the radio to IR and 
X-ray to $\gamma$-ray properties of FRSQ's. Because many sources
in this sample exhibit spectral turnovers near 40 GHz, and these turnovers
indicate that the emission originates from the innermost region of the jet,
they should be
observed with VLBI at 43 GHz (VLBA) and in the 1-3 millimeter range (various arrays) to try to resolve these inner regions
and gain a better understanding of the physical conditions at the base of
the radio jet. In addition, the angular sizes derived from these VLBI images
can be used for more accurate Compton calculations in conjunction with
X-ray and $\gamma$-ray observations. Simultaneous multiwaveband and VLBI
observations would relate the variations at different frequencies both
to each other and to changes in the structure of the jet. Time-dependent
models of SSC emission from relativistic jets should be developed to
replace the crude uniform spherical source approximation used in this
study. Finally, complete samples chosen using various well defined
criteria should be studied thoroughly to establish statistical
relationships as well as to elucidate the effects of those selection
criteria. Together, future studies such as these should lead to a more
complete understanding of the nature of the nonthermal emission in FRSQ's.
\acknowledgments
We thank the anonymous referee for many useful comments and
patience while the final manuscript was being revised.
The authors gratefully acknowledge the use of data from the service
observing programs at both the JCMT and UKIRT, and thank the staffs of
these telescopes for carrying out the observations on our behalf.
The JCMT is operated jointly by the UK Particle Physics and Astronomy
Research Council, the Netherlands Organization for the Advancement of Pure
Research (ZWO), the Canadian National Research Council (NRC), and the
University of Hawaii. UKIRT is operated by the Royal Observatory, Edinburgh
on behalf of the UK Particle Physics and Astronomy Research Council. The
National Radio Astronomy Observatory is operated by Associated Universities,
Inc., under cooperative agreement with the US National Science Foundation.
This material is based upon work supported by the
NSF under grants AST-9116525 (Boston University)
and AST-9120224 (University of Michigan), and by
NASA under grants NAG5-1566, NAG5-1943 (Boston University).
S. Bloom also gratefully acknowledges a grant by the Israel Council for Higher Education during his stay at Technion (Israel Institute of Technology), and
an NRC Research Fellowship while at NASA/GSFC.

\newpage
\centerline{\bf FIGURE CAPTIONS}
\figcaption[fig1.ps]{VLBI images at 8.4 and 22 GHz from the Global VLBI
Network. Parameters of
each map are listed in Table 3. The restoring beam is shown in the
lower left corner of each separate plot.}
\figcaption[fig2.new.ps]{VLBA images at 22 GHz}
\figcaption[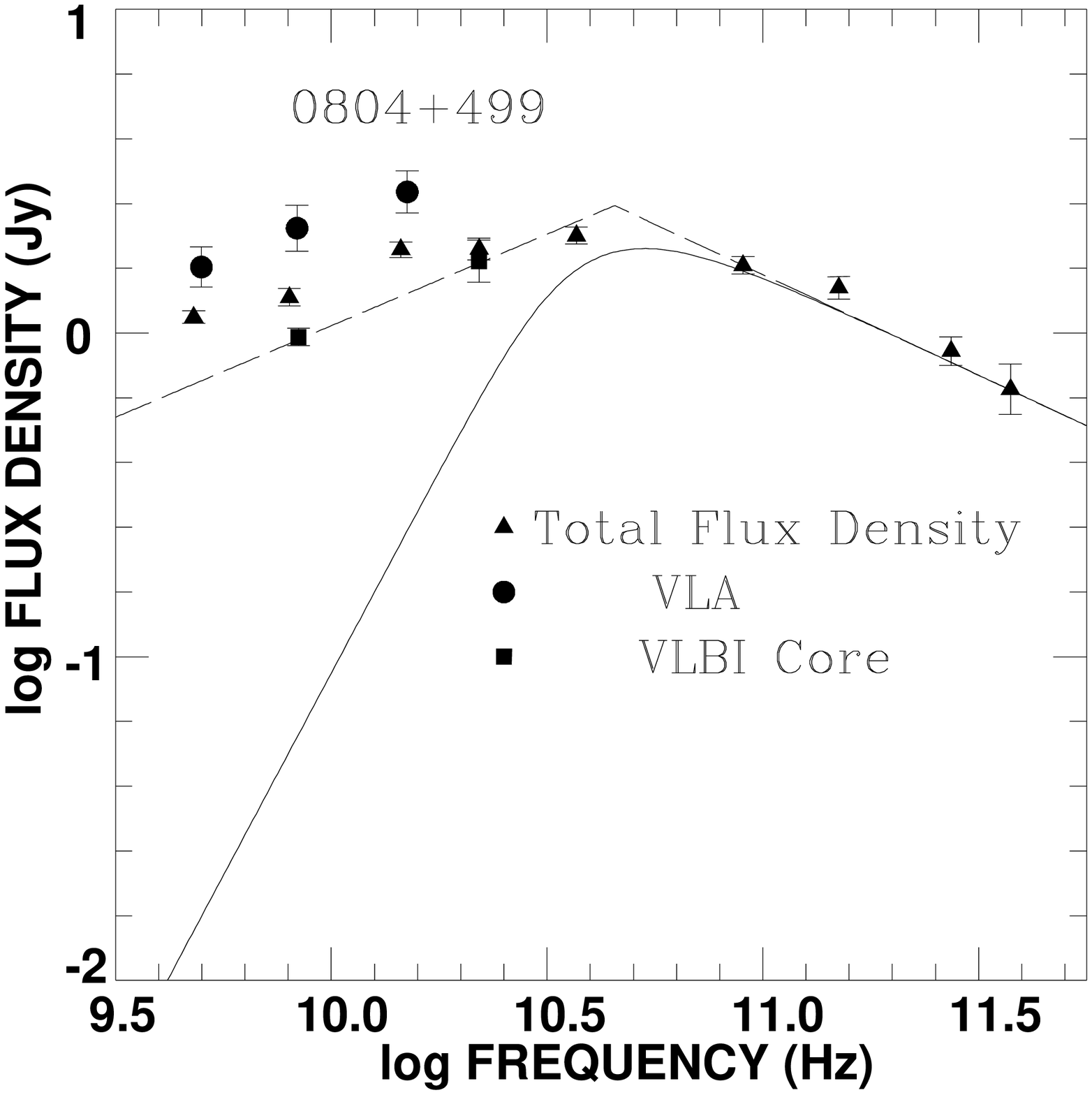]{Determination of the turnover frequency from VLBI and total
flux density data. The dotted lines show the frequency $\nu_n$ at which the
extrapolation of the partially opaque  and optically thick spectra
meet. The solid curve is the synchrotron spectrum for a homogeneous
sphere with spectral turnover at frequency $\nu_m$ as discussed in the
text.}

\figcaption[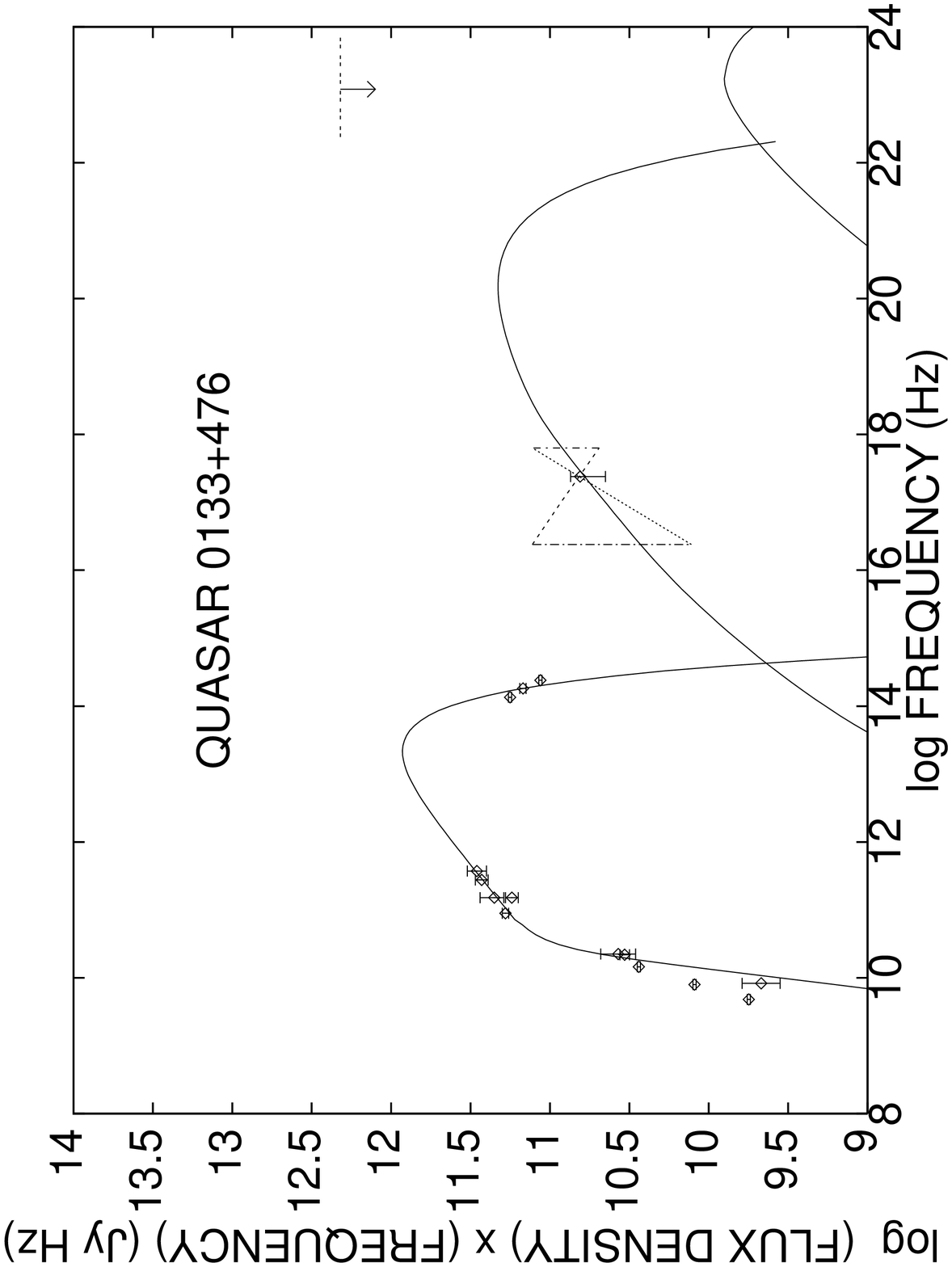]{The synchrotron spectrum (fit by a self-absorbed, uniform
source model) and predicted first order (middle curve)
and second order (rightmost curve) spectra for 0133+476, plotted 
against the contemporaneous data.}

\figcaption[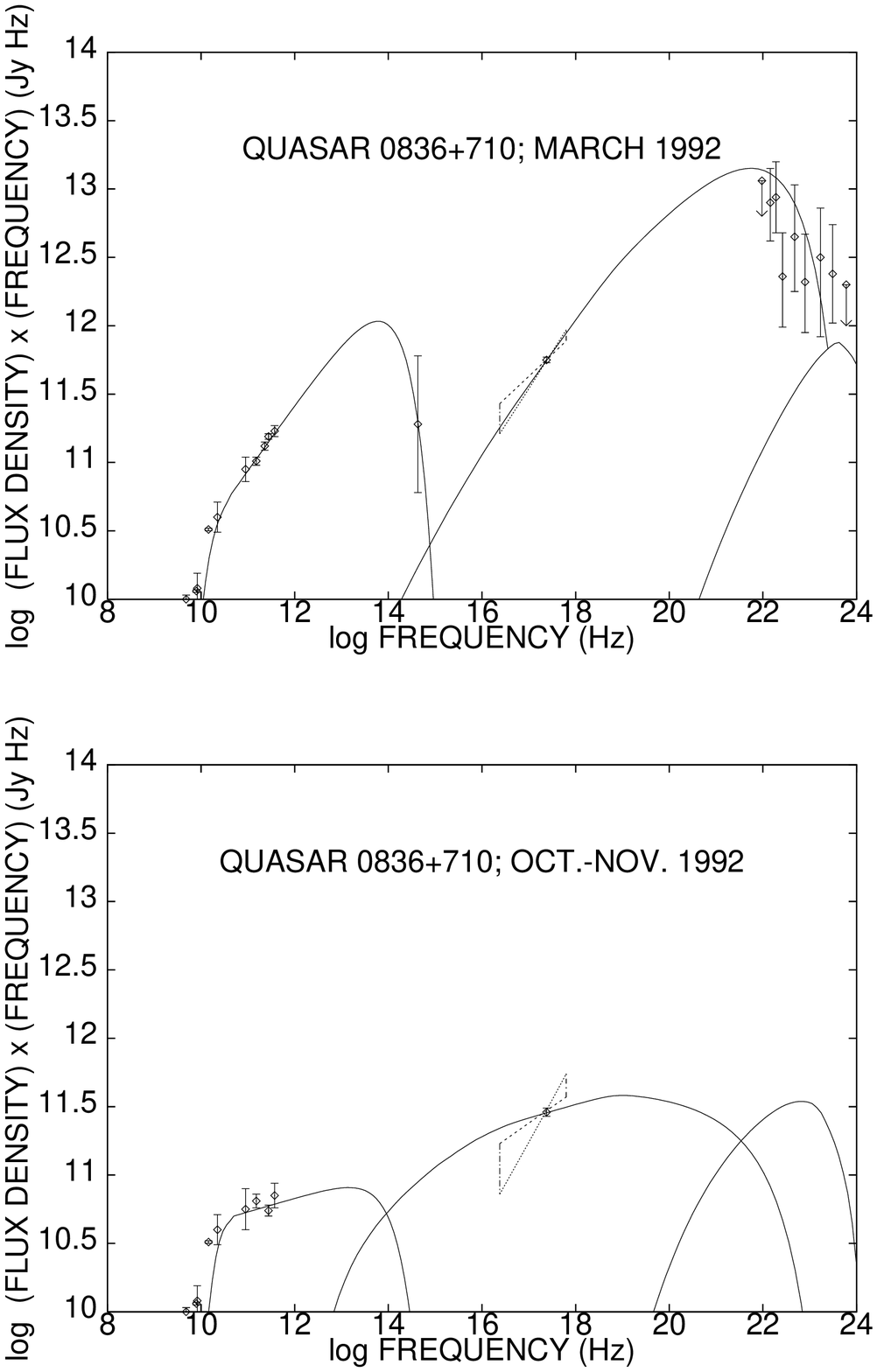]{Similar to Figure 4, except for the quasar 0836+710.
The EGRET gamma-ray data from Thompson et al. (1993)
is plotted to the right on the 1992 March spectrum. Optical data are from
von Linde et al. (1993). The large uncertainty in the optical point
corresponds to the range of fluxes during a short term flare.}

\figcaption[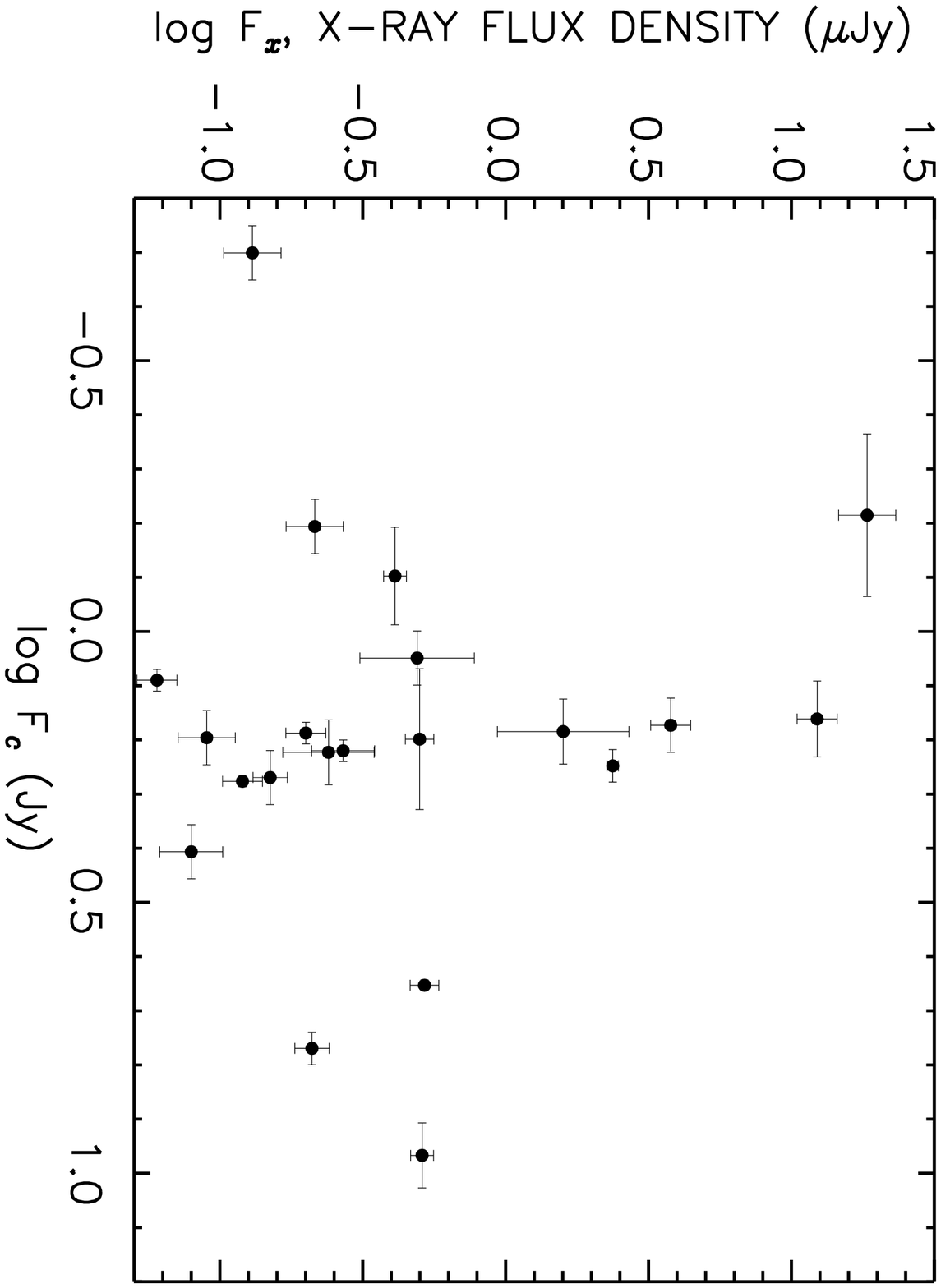]{Correlation plot between VLBI core flux
density at 22 GHz and X-ray flux density at 1 keV.} 

\figcaption[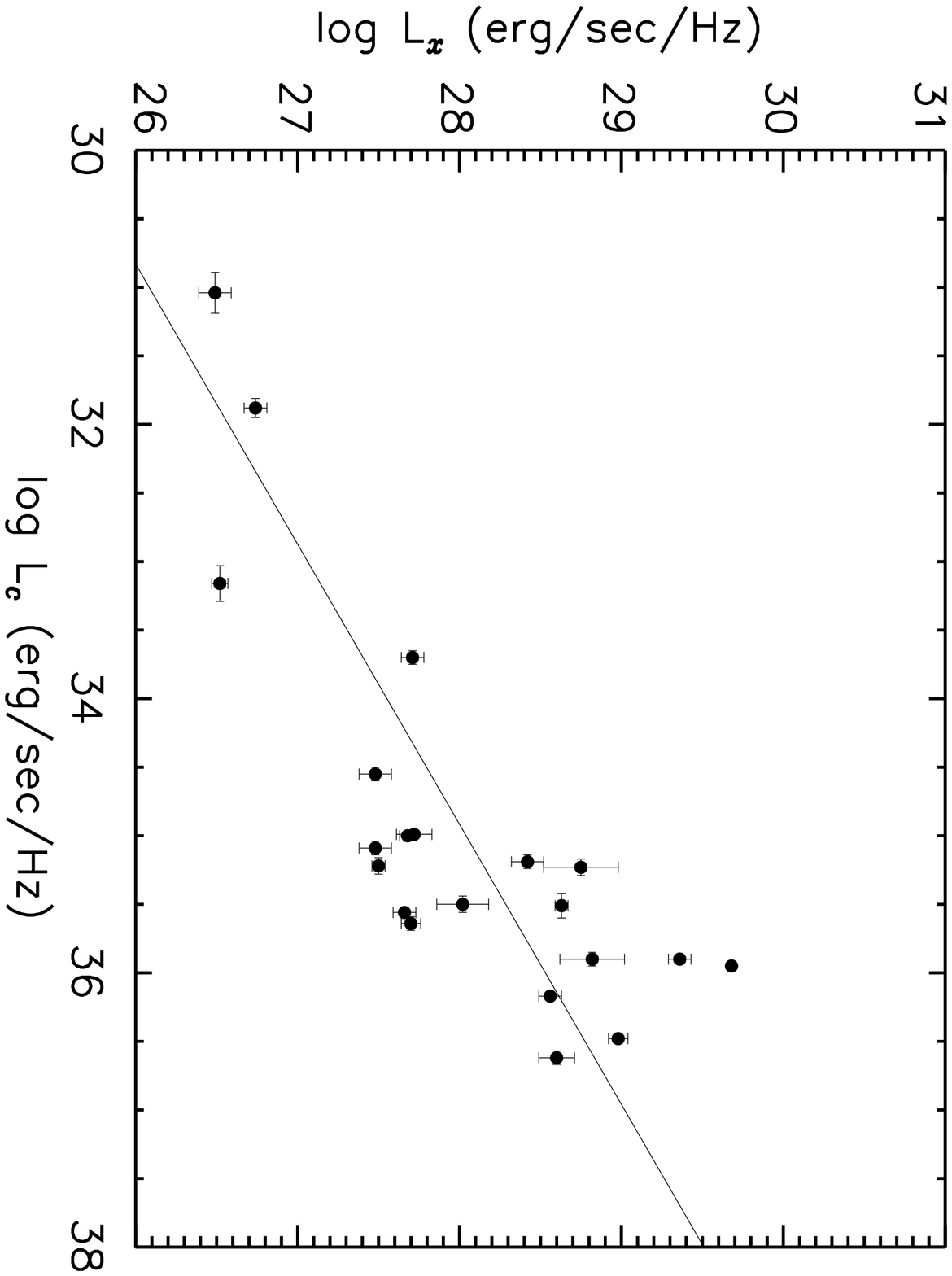]{VLBI core luminosity at 22 GHz
vs. X-ray luminosity at 1 keV. The solid line corresponds to a linear
regression.}

\figcaption[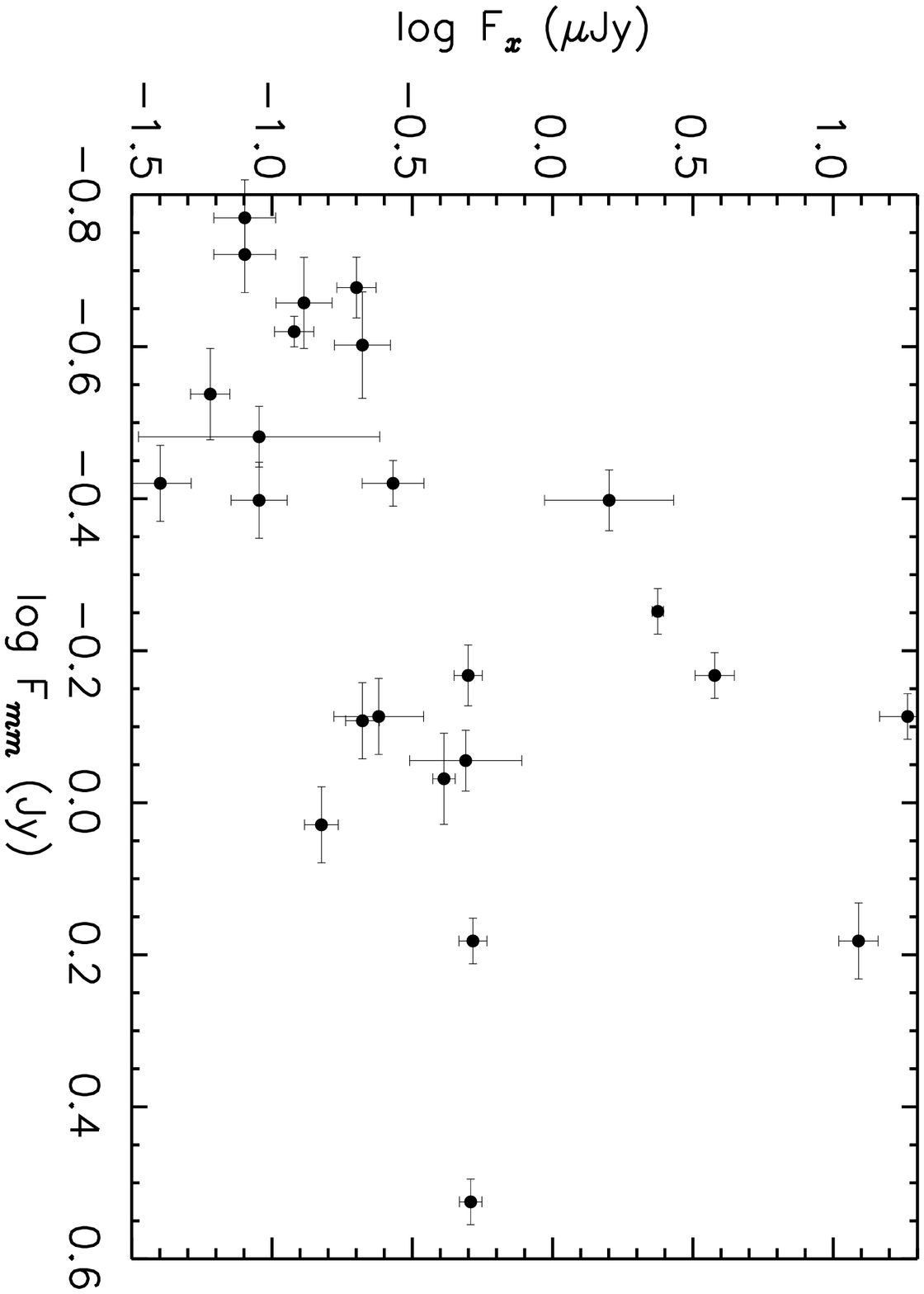]{Millimeter-wave flux density (273 GHz) vs.
X-ray flux density at 1 keV.}

\figcaption[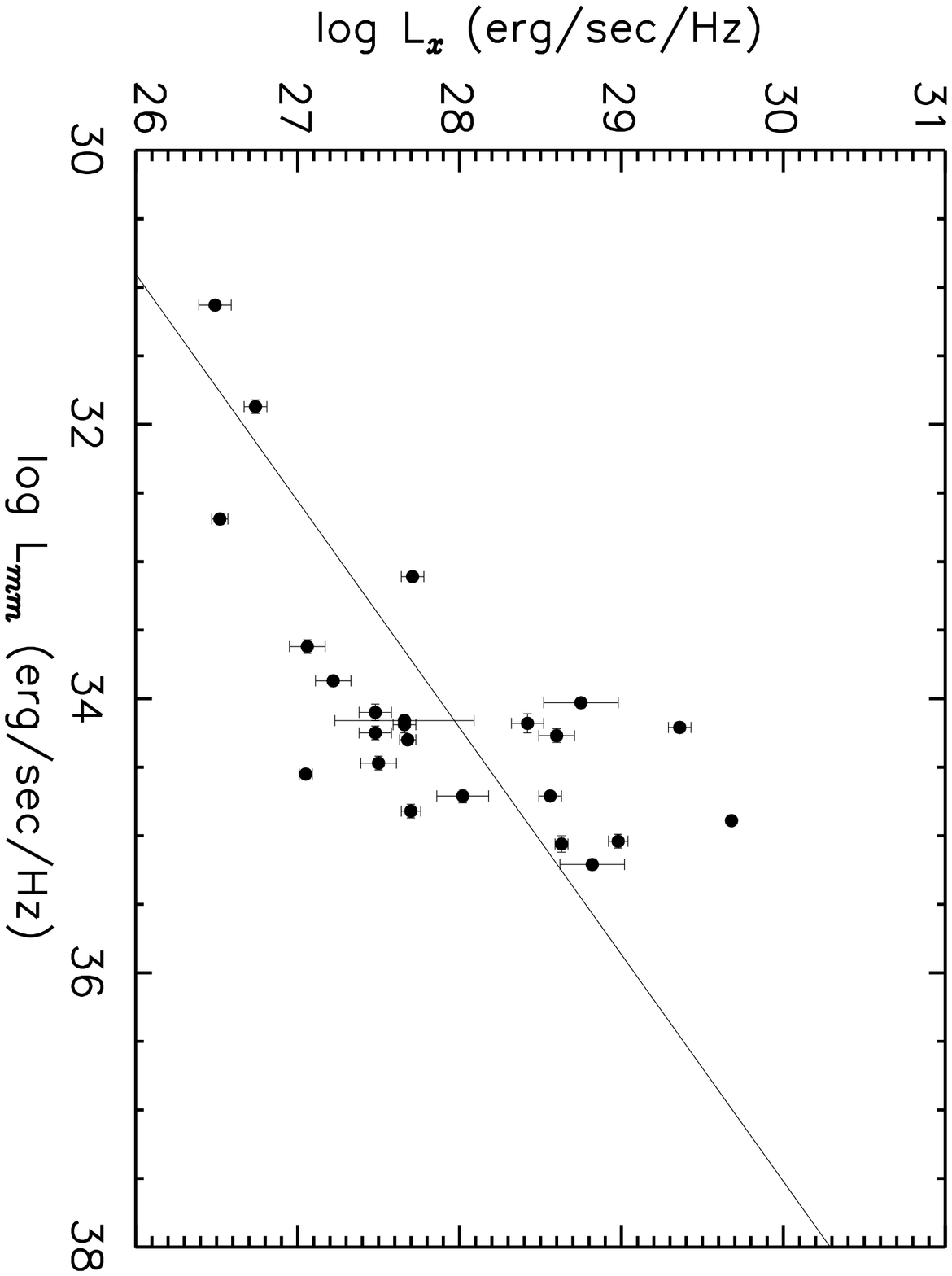]{Millimeter-wave luminosity (273 GHz) vs.
X-ray luminosity at 1 keV.} 

\figcaption[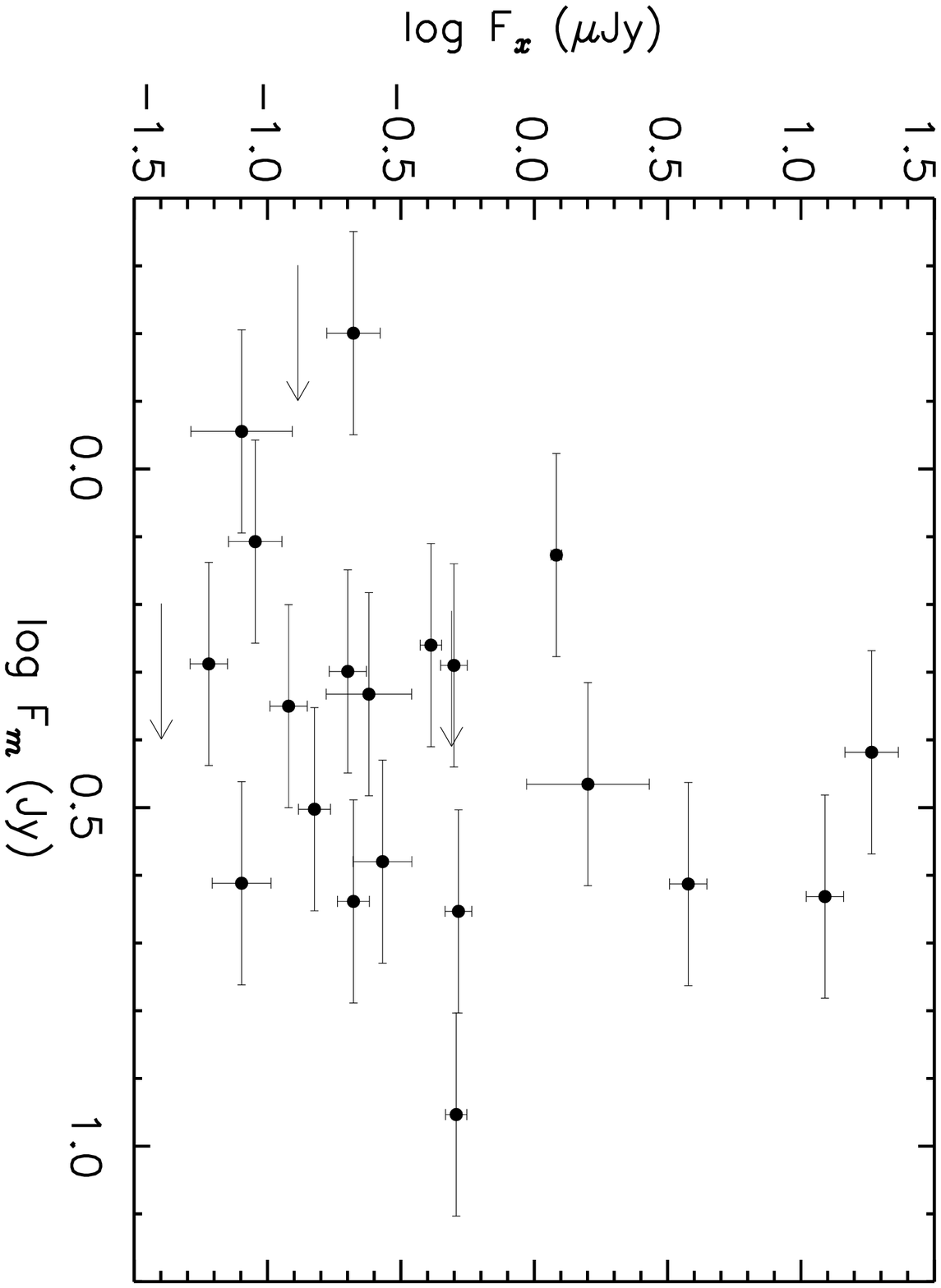]{Flux density at the turnover
frequency $\nu_m$ vs.
X-ray flux density at 1 keV.}

\figcaption[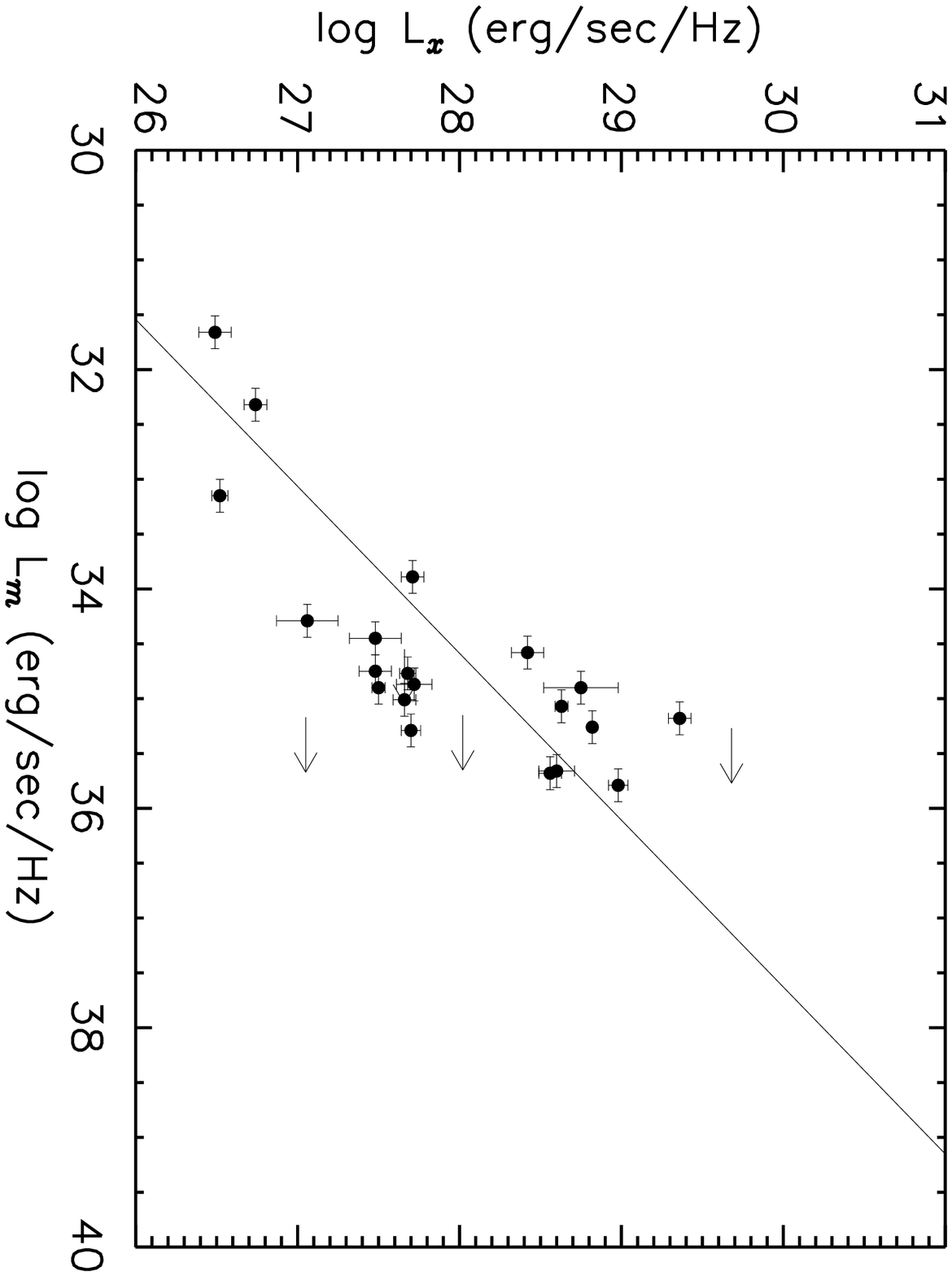]{Luminosity at the turnover
frequency $\nu_m$ vs.
X-ray luminosity at 1 keV. 
The solid line is a linear regression.} 

\figcaption[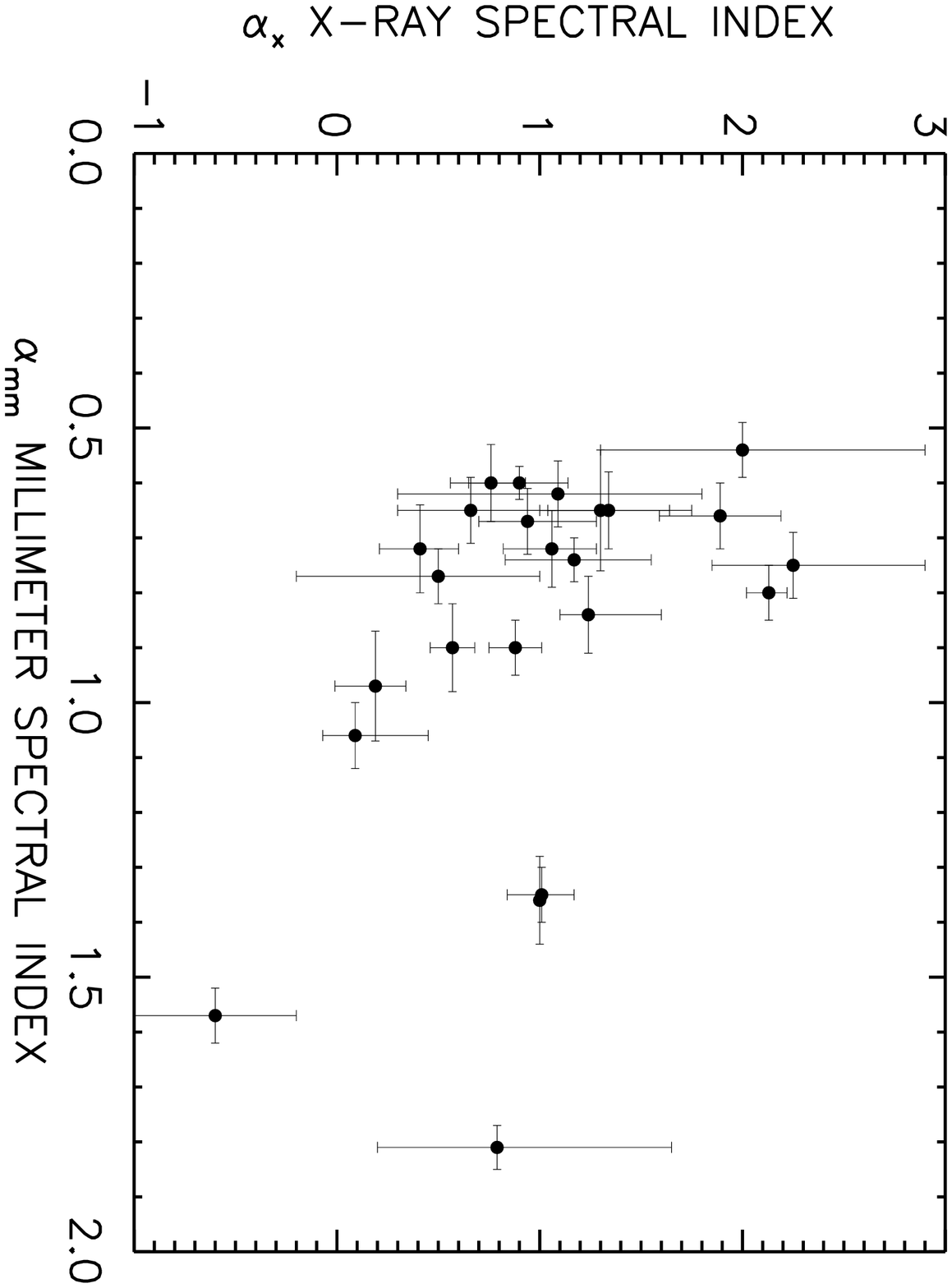]{X-ray spectral index vs. millimeter-wave spectral 
index.}

\figcaption[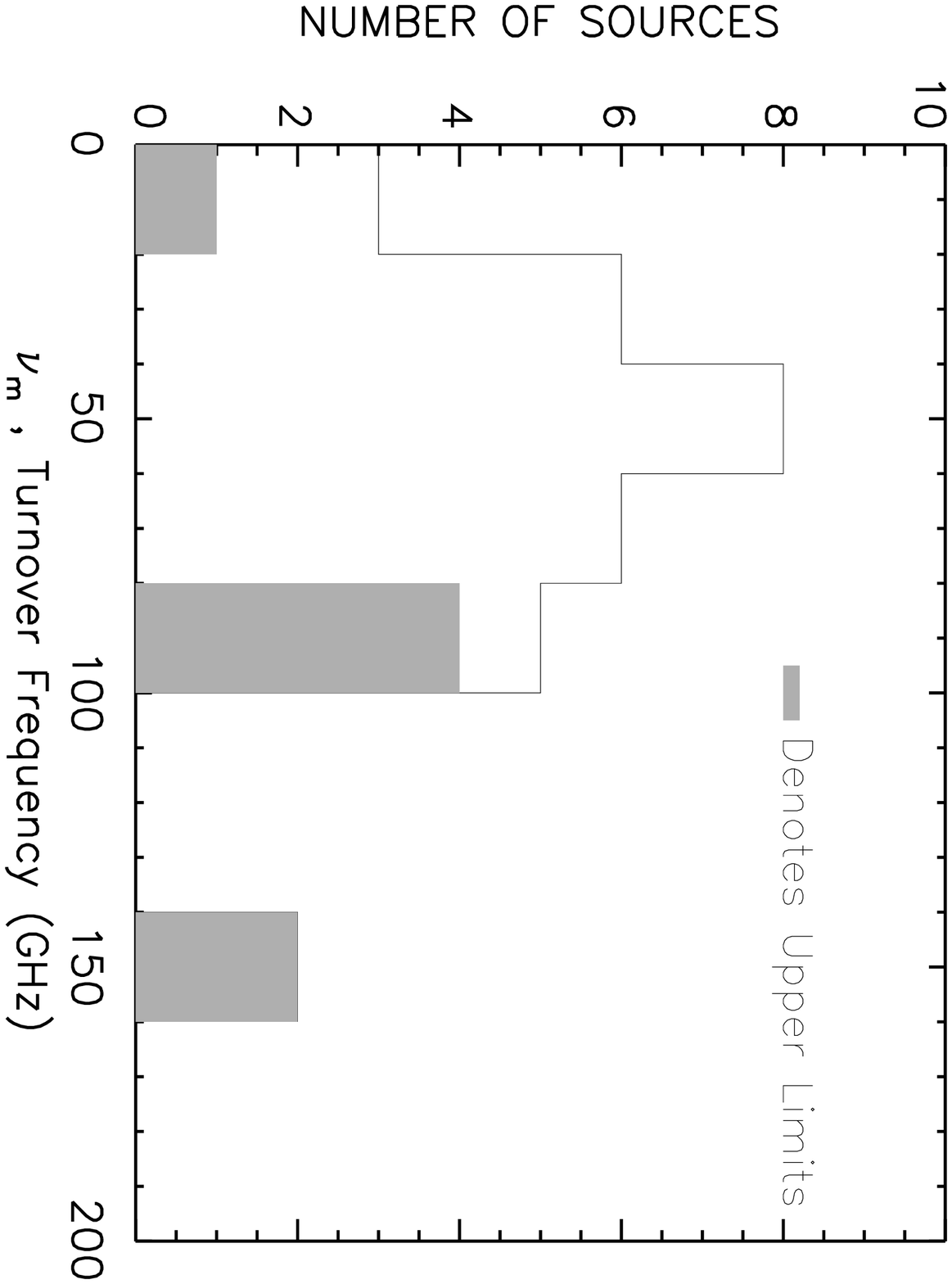]{Histogram of turnover frequencies $\nu_m$.}

\figcaption[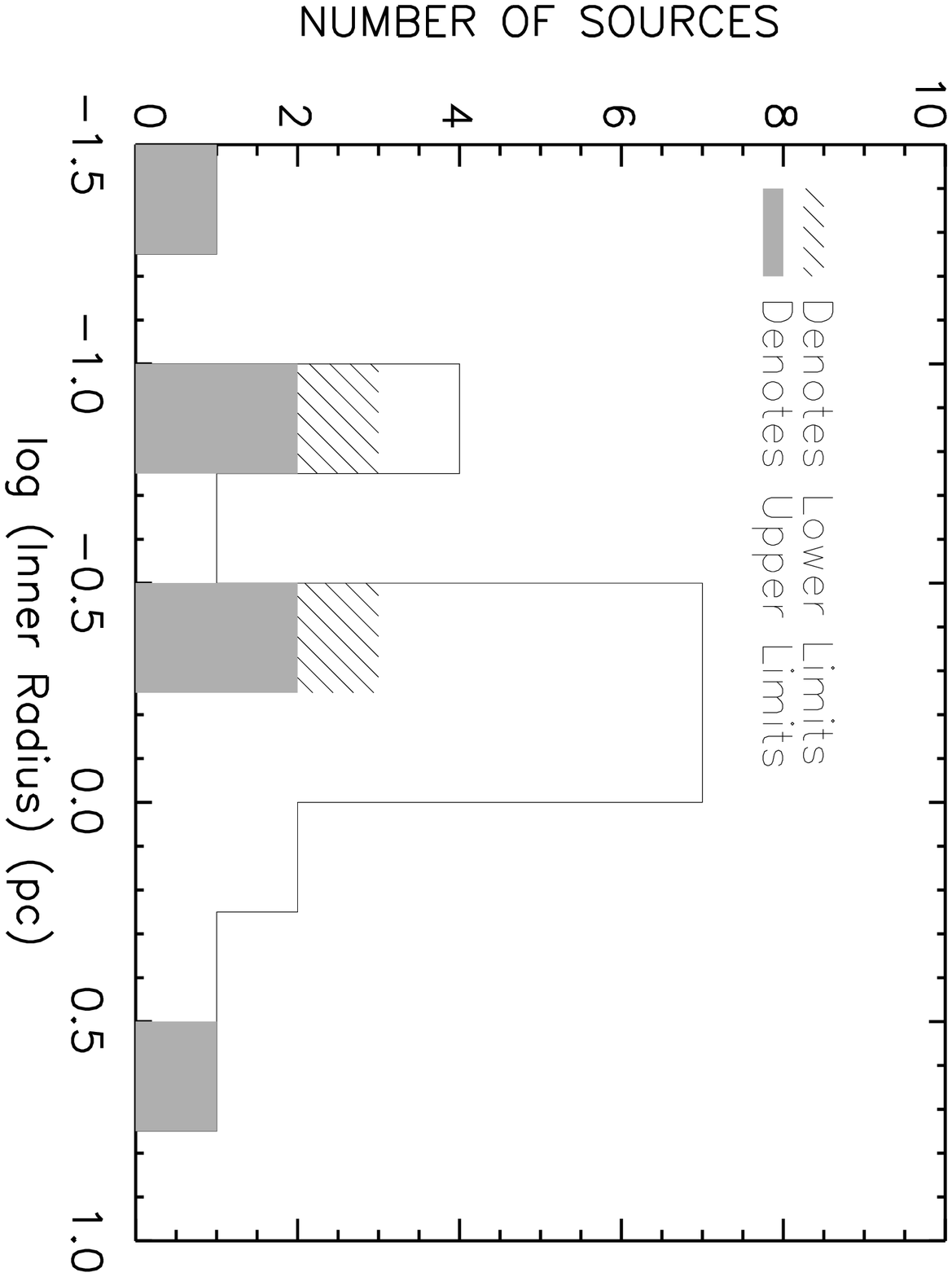]{Histogram of innermost radius measured by our 22GHz
VLBI
observations.}

\figcaption[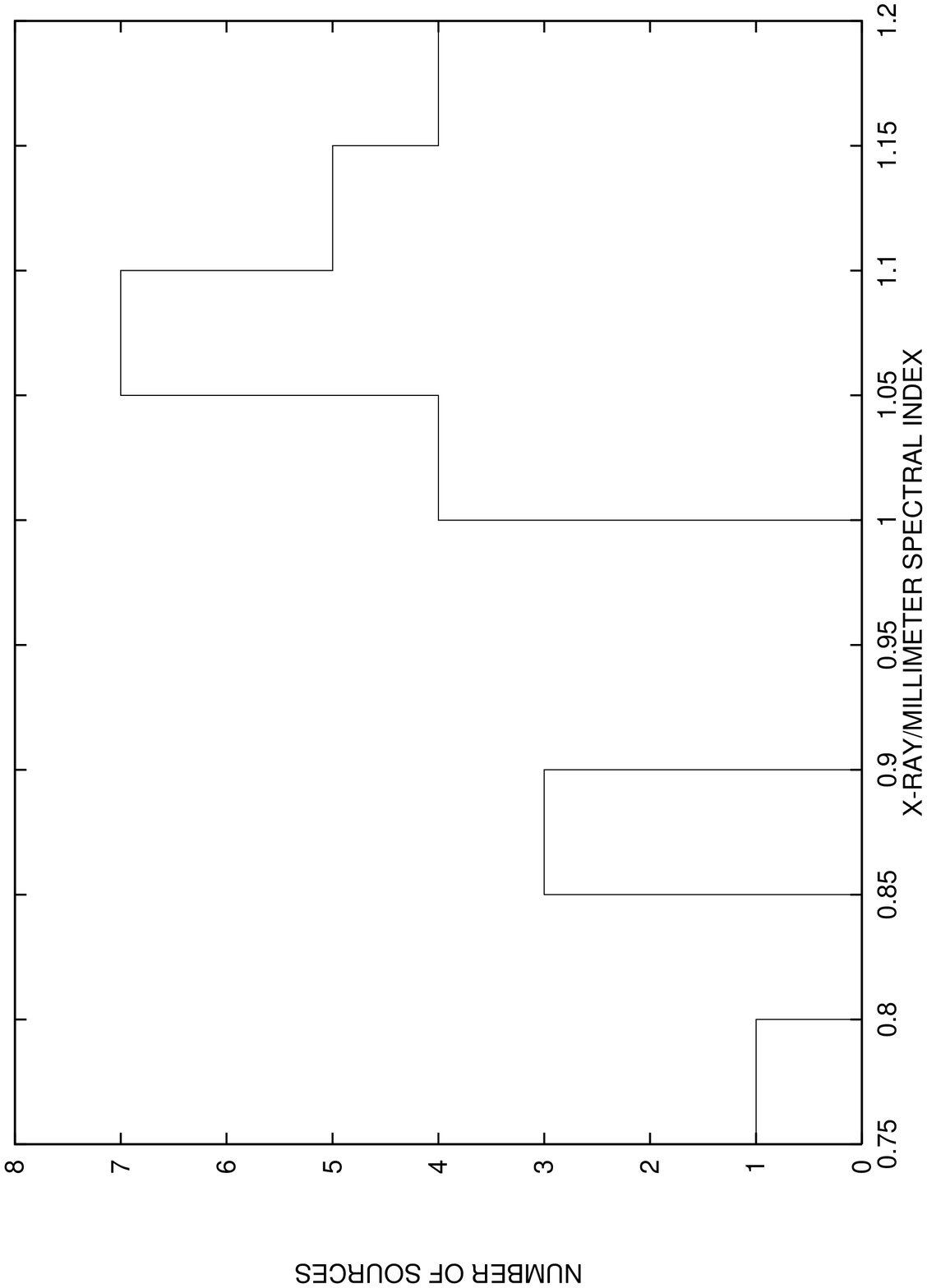]{Histogram of X-ray to millimeter-wave spectral index, defined
as  $\alpha_{mmx}$ = log$(F_x/F_{\rm mm})/$log$(2.4\times 10^{17}/2.7\times 
10^{11})$}

\figcaption[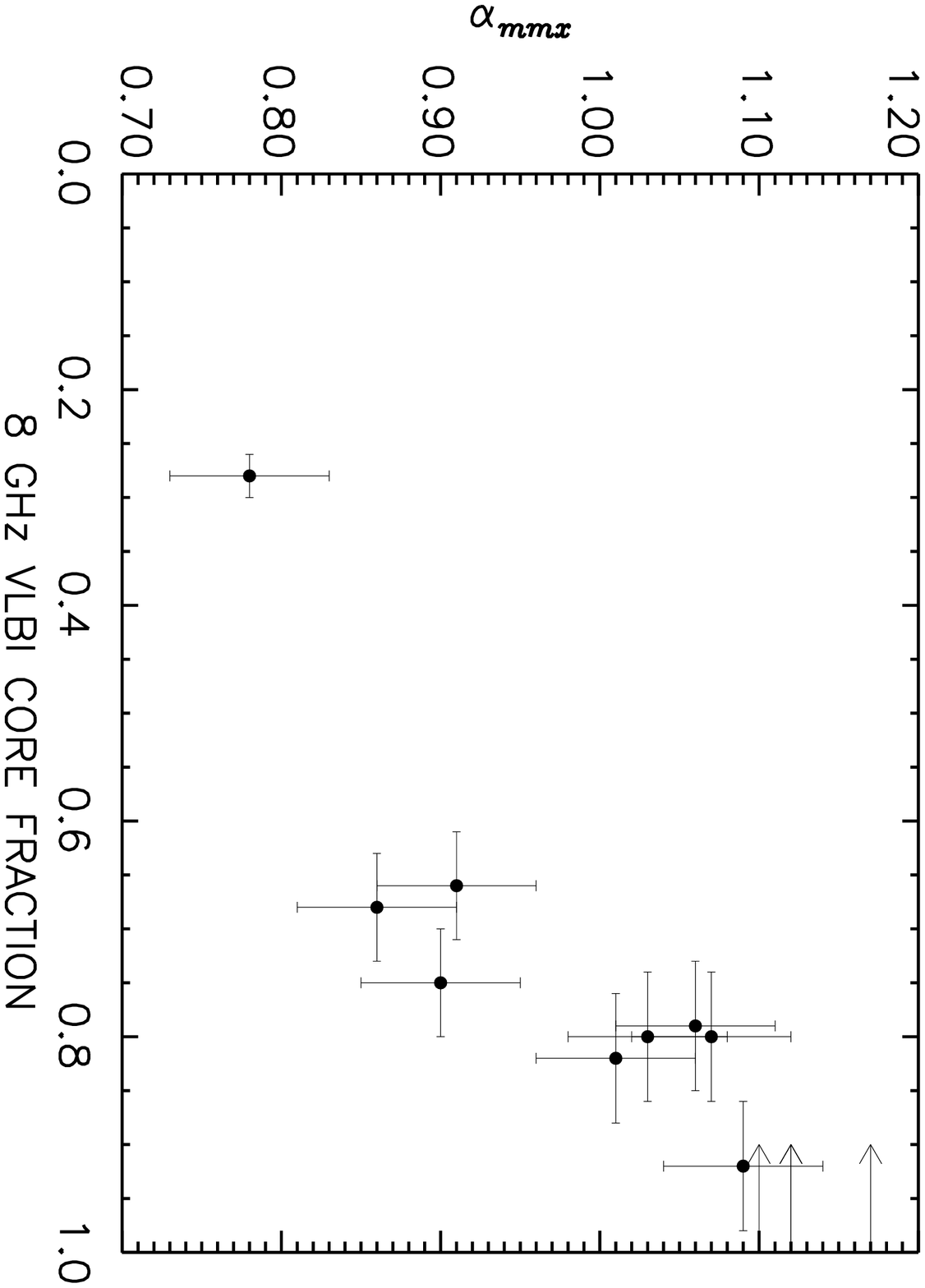]{X-ray to millimeter-wave spectral index $\alpha_{mmx}$ vs.
radio core fraction at 8.4 GHz.}

\figcaption[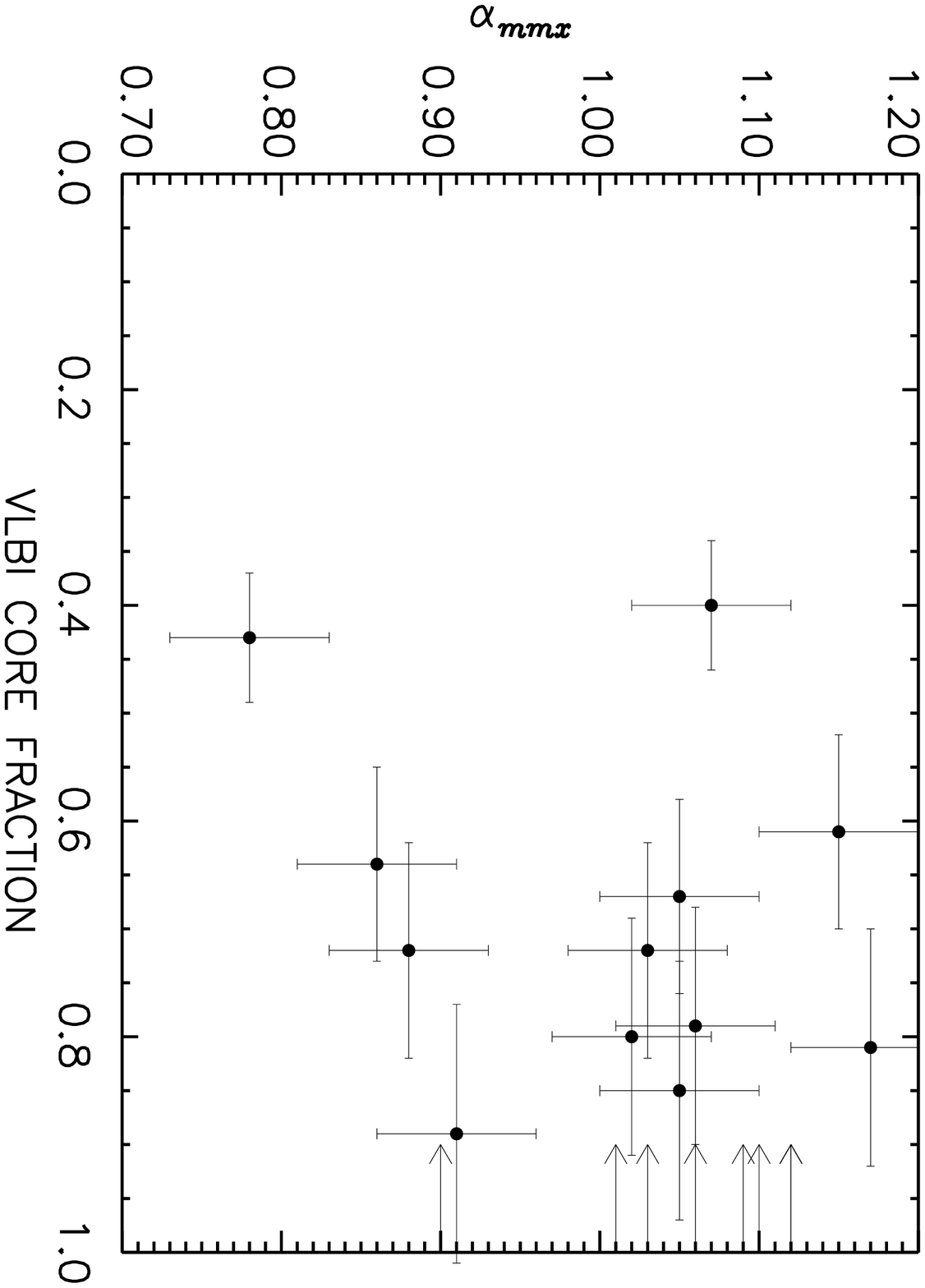]{X-ray to millimeter-wave spectral index $\alpha_{mmx}$ vs.
radio core fraction at 22 GHz.}

\begin{thebibliography}{}

\bibitem[Band \& Grindlay 1986]{ban86} Band, D. L. \& Grindlay, J. E. 1986, Ap J, 308, 576

\bibitem[Bania et al 1991]{ban91} Bania, T. M., Marscher, A. P., \& Barvainis, R. 1991, AJ, 101, 2147

\bibitem[Bevington 1969]{bev69} Bevington, P. R. 1969, Data Reduction and Error Analysis
for the Physical Sciences (New York: McGraw-Hill), 204

\bibitem[Biretta et al 1986]{bir86} Biretta, J. A., et al. 1986, ApJ, 308, 93 

\bibitem[Bloom 1994]{blo94a} Bloom, S. D. 1994, Ph. D. Dissertation, Boston University

\bibitem[Bloom \& Marscher 1991]{blo91} Bloom, S. D. \& Marscher, A. P. 1991, ApJ, 366, 16

\bibitem[Bloom \& Marscher 1996]{blo96} Bloom, S. D. \& Marscher, A. P. 1996, ApJ, 461, 457

\bibitem[Bloom et al. 1994]{blo94b} Bloom, S. D., Marscher, A. P., Gear, W. K., Aller, H., Aller, M., \&  
Ter\"asranta, H. 1994, AJ, 108, 398

\bibitem[Brown et al. 1989]{bro89} Brown, L. M. J., et al. 1989, ApJ, 340, 149.

\bibitem[Browne et al. 1987]{bro87} Browne, I. W. A., \& Murphy, D. W. 1987, MNRAS, 226, 601

\bibitem[Brunner et al. 1994]{bru94} Brunner, H., Lamer, G., Worrall, D. M., \& Staubert, R. 1994,
A\&A, 287, 436

\bibitem[Cohen et al. 1975]{coh75} Cohen, M. H., et al. 1975, ApJ, 201, 249

\bibitem[Comastri et al. 1997]{com97} Comastri, A. et al. 1997, Ap J, 480, 534

\bibitem[Dermer et al. 1993]{der93} Dermer, C. D. 1993, in Compton Gamma-Ray Observatory, ed.
M. Friedlander, N. Gehrels, \& D. Macomb (New York: AIP), 541

\bibitem[Dermer \& Schlickeiser 1993]{ders93} Dermer, C. D. \& Schlickeiser, 
R. 1993, ApJ, 416, 458 

\bibitem[Eckart et al. 1986]{eck86}Eckart, A., et al. 1986, A\&A, 168, 17

\bibitem[Eckart et al. 1987]{eck87} Eckart, A., et al. 1987, A\&AS, 67, 121

\bibitem[Elvis et al. 1989]{elv89} Elvis, M., Lockman, F. J., \& Wilkes, B. J. 1989, AJ, 97, 777

\bibitem[Fichtel et al. 1994]{fic94} Fichtel, C. E., et al. 1994, ApJS, 94, 551

\bibitem[Frail et al. 1994]{fra94} Frail, D. A., Weisberg, J. M., Cordes, J. M., \& Mathers, C. 1994,
ApJ, 436, 144

\bibitem[Gear et al. 1994]{gea94}Gear, W. K., et al. 1994, MNRAS, 267, 167

\bibitem[Ghisselini et al. 1985]{ghi85}Ghisselini, G., Maraschi, L., \& Treves, A. 1985, A\&A, 146, 204

\bibitem[Grandi et al. 1997]{gra97}Grandi, P. et al. 1997, Ap J, 487, 636

\bibitem[Hartman et al. 1992]{har92}Hartman, R. C., et al. 1992, in The Compton Observatory
Science Workshop, ed. C. R. Shrader, N. Gehrels, \& B. Dennis (NASA Conf. 
Publ 3137), 116
%

\bibitem[Isobe et al. 1986]{iso86} Isobe, T., Feigelson, E. D., \& Nelson, P. I. 1986, ApJ, 306, 490

\bibitem[Isobe \& Feigelson 1990]{iso90} Isobe, T. \& Feigelson, E. D.  1990, BAAS, 22, 917

\bibitem[Jones et al. 1974a]{jon74a} Jones, T. W., O'Dell, S. L., \& Stein, W. A. 1974a, ApJ, 188, 353

\bibitem[Jones et al. 1974b]{jon74b} Jones, T. W., O'Dell, S. L., \& Stein, W. A. 1974b, ApJ, 192, 261

\bibitem[Kembhavi et al. 1986]{kem86} Kembhavi, A., Feigelson, E. D., \& Singh, K. P. 1986, MNRAS, 220, 51

\bibitem[K\"onigl 1981]{kon81} K\"onigl, A. 1981, ApJ, 243, 700

\bibitem[Krichbaum et al. 1990]{kri90} Krichbaum, T. P., et al. 1990, A\&A, 230, 271

\bibitem[Ku et al. 1980]{kuw80} Ku, W. H.-M., Helfand, D. J., \& Lucy, L. B. 1980, Nature, 288, 323

\bibitem[Leach et al. 1995]{lea95} Leach, C. M., Mc Hardy, I. M., \& Papadakis, I. E. 1995, MNRAS, 272, 221

\bibitem[Ledden et al. 1985]{led85} Ledden, J. E., \& O'Dell, S. L. 1985, ApJ, 298, 630

\bibitem[Liszt et al. 1993]{lis93} Liszt, H. S., \& Wilson, R. W. 1993, ApJ, 403, 663

\bibitem[Malaguti et al. 1994]{mal94} Malaguti, G., Bassani, L., \& Caroli, E. 1994, ApJS, 94, 517

\bibitem[Marscher 1977]{mar77} Marscher, A. P. 1977, ApJ, 216, 244

\bibitem[Marscher 1980]{mar80} Marscher, A.P. 1980, ApJ, 235, 386

\bibitem[Marscher 1987]{mar87} Marscher, A. P. 1987 in Superliminal Radio Sources, ed. 
J. A. Zensus \& T. J. Pearson (Cambridge Univ. Press) 280

\bibitem[Marscher 1988]{mar88} Marscher, A. P. 1988, ApJ, 334, 552

\bibitem[Marscher \& Gear 1985]{mar85} Marscher, A. P., \& Gear, W. K. 1985, ApJ, 298, 114

\bibitem[Marscher et al. 1992]{mar92} Marscher, A. P., Gear, W. K., \& Travis, J. P. 1992, in Variability
of Blazars, ed. E. Valtaoja \& M. Valtonen (Cambridge Univ. Press), 85

\bibitem[Marscher et al. 1991]{mar91} Marscher, A. P., Zhang, Y. F., Shaffer, D. B., Aller, H. D., \&
Aller, M. F. 1991, ApJ, 371, 491

\bibitem[Mattox et al. 1993]{mat93} Mattox, J. R., et al. 1993, ApJ, 410, 609

\bibitem[Mattox et al. 1996]{mat96} Mattox, J. R., et al. 1996, ApJ, 461, 396

\bibitem[Melia \& K\"onigl]{mel89} Melia, F., \& K\"onigl, A. 1989, ApJ, 340, 162

\bibitem[Owen et al. 1981]{owe81} Owen, F. N., Helfand, D. J., \& Spangler, S. R. 1981, ApJ, 250, L55

\bibitem[Padovani 1992]{pad92} Padovani, P. 1992, A\&A, 256, 399

\bibitem[Pearson \& Readhead 1984]{pea84} Pearson, T. J. \& Readhead, A. C. S. 1984, ARA\&A, 22, 97

\bibitem[Pearson \& Readhead 1988]{pea88} Pearson, T. J. \& Readhead, A. C. S. 1988, ApJ, 328, 114

\bibitem[Petre et al. 1984]{pet84} Petre, R., Mushotzsky, R. F., Krolik, J. H., \& Holt, S. S. 1984, ApJ, 280, 499

\bibitem[Sambruna 1997]{sam97} Sambruna, R. 1997, Ap J, 487, 536 

\bibitem[Schwab \& Cotton 1983]{sch83} Schwab, F. R. \& Cotton, W. D. 1983, AJ, 88, 688

\bibitem[Stark et al. 1992]{sta92} Stark, A. A., et al. 1992, ApJS, 79, 77

\bibitem[Steppe et al. 1988]{ste88} Steppe, H., et al. 1988, A\&AS, 75, 317

\bibitem[Swanenburg et al. 1978]{swa78} Swanenburg, B. N., et al. 1978, Nature, 275, 298

\bibitem[Tananbaum et al. 1979]{tan79} Tananbaum, H., et al. 1979, ApJ, 234, L9

\bibitem[Thompson et al. 1993]{tho93} Thompson, D. J., et al. 1993, ApJ, 415, L13

\bibitem[Thompson et al. 1995]{tho95} Thompson, D. J. , et al 1995. ApJS, 101, 
259

\bibitem[Thompson et al. 1996]{tho96} Thompson, D. J. , et al 1996, ApJS, 107, 
227

\bibitem[Tornikoski et al. 1994]{tor94} Tornikoski, M., et al. 1994, A\&A, 289, 673

\bibitem[Unwin et al. 1994]{unw94} Unwin, S. C., et al. 1994, ApJ, 432, 103

\bibitem[Unwin et al. 1997]{unw97} Unwin, S. C., et al. 1997, ApJ, 480, 596

\bibitem[von Linde  et al. 1993]{von93} von Linde, J., et al. 1993, A\&A, 267, L23

\bibitem[von Montigny et al. 1995]{von95} von Montigny, C., et al. 1995, ApJ, 440, 525

\bibitem[Wehrle \& Mattox 1994]{weh94} Wehrle, A. E., \& Mattox, J. R. 1994, in The Second Compton
Symposium, ed. C. E. Fichtel, N. Gehrels, \& J. P. Norris, AIP
Conf. Proc. 304, 688

\bibitem[Wilkes 1992]{wil92} Wilkes, B. J. 1992, ROSAT--PROS Users Guide (Cambridge, MA: Smithsonian
Astrophys. Obs.)

\bibitem[Wilkes \& Elvis 1987]{wil87} Wilkes, B. J., \& Elvis, M. 1987, ApJ, 323, 243

\bibitem[Worrall et al. 1987]{wor87} Worrall, D. M., Giommi, P., Tananbaum, H., \& Zamorani, G.  1987, ApJ, 
313, 596

\bibitem[Worrall \& Wilkes 1990]{wor90} Worrall, D. M. and Wilkes, B. J. 1990, ApJ, 360, 396

\bibitem[Zamorani et al. 1981]{zam81}Zamorani, G. et al. 1981, ApJ, 245, 357

\bibitem[Zhang 1994]{zha94a} Zhang, Y. F. 1994, PhD Dissertation, Boston University

\bibitem[Zhang \& Marscher 1994]{zha94b} Zhang, Y. F., \& Marscher, A. P. 1994, in Proceedings of the
1993 ROSAT Science Symposium, ed. E. Schlegel \& R. Petre (New York:
AIP)
\bibitem[Zhou et al. 1997]{zho97} Zhou, Y. Y. et al 1997, ApJ, 484, L47
\end{thebibliography}
\end{document}


\begin{deluxetable}{lllllll}
\pagestyle{empty}
\tablecolumns{7}
\tablewidth{0pt}
\tablenum{1}
\tablecaption{Summary of Observations}
\tablehead{\colhead{Source}&\colhead{z}&\colhead{VLBI}&\colhead{JCMT}&
\colhead{UKIRT}&\colhead{ROSAT}&\colhead{EGRET}
\nl
\colhead{(1)}&\colhead{(2)}&\colhead{(3)}&\colhead{(4)}&\colhead{(5)}
&\colhead{(6)}& \colhead{(7)}}
\startdata
0016+731&1.76&Jun 1992& Aug 1991&\nodata&Jul 1992&Jul/Aug 1992\nl
        &    &Sep 1992& Oct 1992&\nodata&\nodata&\nodata\nl
0133+476&0.86&Jun 1992& Aug 1991& Aug 1992& Aug 1992& Nov/Dec 1991\nl
        &    &Sep 1992& Oct 1992&\nodata&\nodata&\nodata\nl
0212+735&2.37&Nov 1992& Aug 1991&\nodata& Feb 1991& Jul/Aug 1992\nl
        &    &Mar 1993& \nodata &\nodata& \nodata& \nodata\nl
NRAO 140&1.26&Jun 1991& Aug 1991& Aug 1992 &Aug 1992& May 1991\nl
        &     &Nov 1991& Oct 1992& \nodata  &\nodata & \nodata\nl
3C 111  &0.05&Jun 1991 & Aug 1991& \nodata& Mar 1991& Nov/Dec 1991\nl
        &    &Nov 1991 & Oct 1992& \nodata& Feb 1993& \nodata \nl
0429+415&0.41&Jun 1992 & Aug 1991& \nodata& \nodata& \nodata \nl
        &    &Sep 1992 & Oct 1992& \nodata& \nodata& \nodata \nl
3C 120  &0.03&Jun 1991 & Aug 1991& \nodata& Feb 1991& Aug 1992\nl
        &    &Nov 1991 & Oct 1992& \nodata& \nodata & \nodata \nl
0552+398&2.36&Jun 1992 & Aug 1991& \nodata&Mar 1993&June 1992\nl
        &    &Sep 1992 & Oct 1992& \nodata&\nodata &\nodata\nl
0642+449&3.40&Jun 1991 & Aug 1991& \nodata&Mar 1991& \nodata\nl
        &    &\nodata  &Jan  1992& \nodata &\nodata &\nodata\nl
0736+017&0.19&Jun 1991 & Aug 1991 & \nodata& Apr 1991& Nov 1992\nl
        &    &\nodata  & Oct 1992 & \nodata& Apr 1992& \nodata\nl
0804+499&1.43&Jun 1992 & Aug 1991 & Nov 1993& Oct 1992& Jan 1992\nl
        &    &Sep 1992 & Oct 1992 & Mar 1994& \nodata & \nodata \nl
0820+560&1.42&\nodata  & Aug 1991 & Dec 1993& Oct 1993& \nodata \nl
0836+710&2.17&Jun 1992& Mar 1992 & \nodata & Mar 1992 & Jan 1992\nl
        &    &Sep 1992& Oct 1992 & \nodata & Nov 1992 & Mar 1992\nl
0917+449&2.18&Jan 1995& \nodata  & \nodata & Oct/Nov 1992&\nodata\nl
4C 39.25&0.70&\nodata& Aug 1991 & Nov 1993 &Apr 1991& Sep/Oct 1992\nl
        &    &\nodata& Oct 1992 & \nodata & Apr 1993& \nodata\nl   
0945+408&1.25&Nov 1992&Oct 1992& May 1994& Nov 1993& Sept/Oct 1992\nl
        &    &Mar 1993&\nodata & May 1994& Nov 1993& Sep/Oct 1993\nl
        &    &Jan 1995&\nodata &\nodata  &\nodata  & \nodata\nl
0954+556&0.91&\nodata& Oct 1992& Dec 1993& Nov 1993& Jun/Jul 1991\nl
0955+476&1.87&Jan 1995& Oct 1992& Dec 1993& Nov 1993& \nodata \nl
1150+497&0.334&\nodata&Mar 1992& Jun 1992& \nodata &Jun/Jul 1992\nl
        &     &\nodata&Oct 1992&\nodata  & \nodata &\nodata\nl
        &     &Jan 1995&\nodata&\nodata  & \nodata &\nodata\nl
1150+812&1.25 &\nodata&Mar 1992&\nodata  & Apr 1992&Jan 1991\nl
        &     &\nodata&\nodata &\nodata  & May 1992 &\nodata\nl
        &     &\nodata&\nodata & \nodata & Oct 1992 &\nodata\nl
1611+343&1.40 & Nov 1992&Oct 1992&Aug 1992&Sep 1992&Sep 1992\nl
        &     & \nodata& \nodata&\nodata  &Jan 1993&\nodata\nl
        &     &May 1996 &\nodata &\nodata &\nodata & \nodata\nl
1633+382&1.81 & Sep 1992&Oct 1992&Aug 1992& Aug 1992&Nov 1992\nl
        &     & Mar 1993&\nodata &\nodata &\nodata  &Sep 1993\nl
1638+398&1.66 &\nodata  & Oct 1992&Sep 1992& Sep 1991&\nodata\nl
1641+399&0.60 &Nov 1992 & Oct 1992&\nodata & Jun 1990&Sep 1991\nl
        &     &Mar 1993 & \nodata &\nodata & Jan 1992&\nodata\nl
1828+487&0.69 &\nodata  &Aug 1991 & \nodata&\nodata & Sep 1991\nl
1954+513&1.22 &Jan 1995&Aug 1991 &\nodata &\nodata & May/Jun 1991\nl
2005+403&1.74 &Jan 1995&Aug 1991 &\nodata &\nodata &May/Jun 1991\nl
2037+511&1.69 &\nodata  &Aug 1991 &\nodata &\nodata &May/Jun 1991\nl
        &     &\nodata  &Oct 1992 &\nodata &\nodata &\nodata \nl
        &     & Jan 1995&\nodata  &\nodata &\nodata &\nodata \nl
2136+141&2.43 &Jun 1991 &Aug 1991 &\nodata &May 1991 &Jan/Feb 1992\nl
        &     &Nov 1991 &\nodata  &\nodata&\nodata &\nodata \nl
2201+315&0.30 &Jan 1995 &Aug 1991 &\nodata&Dec 1993& May/Jun 1991\nl
\enddata
\end{deluxetable}